%% file: main.tex
\documentclass[12pt,notitlepage]{article}
\usepackage{amsfonts}
\usepackage{amssymb}
\usepackage{graphicx}
\usepackage{amsmath}
\usepackage{endnotes}
\usepackage{setspace}
\usepackage{setspace}
\usepackage{fancyvrb}
\usepackage{tikz}

\newcommand{\bb}{\bar{\beta}}
\newcommand{\sT}{s_{\mathcal{T}}}

\newcommand{\dt}{{\frac{\partial}{\partial t}}}

\usepackage{hyperref}
\usepackage{amsmath}

\begin{document}

\title{Addressing the Herd Immunity Paradox Using\\
    Symmetry, Convexity Adjustments and Bond Prices
            }
\author{Peter Cotton\footnote{Intech Investments. We thank Maria Ma for helpful feedback on the manuscript.}}
\date{Apr 26, 2020}
\maketitle

\begin{abstract}
In constant parameter compartmental models an early onset of herd immunity is at odds with estimates of R values from early stage growth. This paper utilizes a result from the theory of interest rate modeling, namely a bond pricing formula of Vasicek, and an approach inspired by a foundational result in statistics, de Finetti's Theorem, to show how the modeling discrepancy can be explained. Moreover the difference between predictions of classic constant parameter epidemiological models and those with variation and stochastic evolution can be reduced to simple ``convexity'' formulas. A novel feature of this approach is that we do not attempt to locate a true model but only a model that is equivalent after permutations.
Convexity adjustments can also be used for cross sectional comparisons and we derive easy to use rules of thumb for estimating threshold infection level in one region given knowledge of threshold infection in another. 
\end{abstract}

\include{overview}

\include{definetti}

\include{convexity}

\include{vasicek}

\section{Summary}

It is possible that some compartmental models for disease spread suffer from a pronounced flaw of averages, and it is notable that in comparing model predictions for early stage growth and peak infection, these errors combine as we have shown in Section \ref{sec:convexity}. However we need not discard the elegance and simplicity of compartmental models, if simple corrections for variation are available.   

We have treated the predictions of models exhibiting variation as perturbations of models that do not. This follows a long tradition in applied mathematics and in particular homogenization \cite{Nolen2008AProblems}. It helps us build on the intuition of the simpler models while still capturing the first order effects. 

We have shown that mixture models and stochastic models are not only interesting in their own right, and may resolve serious model mispecification issues, but also find interpretation within an approach that is motivated by de Finetti's Theorem. In this setting, mixture models with tractable convexity adjustments can help locate a model that is equivalent to nature's, at least as far as symmetric aggregate quantities are concerned.  

In this approach it is not a requirement that the amount of variation implied by convexity adjustments be plausible purely from the point of view of population variation. And it may be possible to provide actionable intelligence without ever coming close to the true model. 

From an analytical perspective it is clear from the size of the adjustments that care is required any time we have variation that reflects genuine heterogeneity. This is true of cross sectional comparisons also, if the amount of variation can be expected to vary. Population density may have only a weak relationship to contagion but because it varies so dramatically a convexity adjustment may be warranted. 

Finally, we have appealed to the usefulness of growth models of a different kind, taken from the interest rate literature. Though this connection is limited to early stage growth the flexibility provided may be of benefit to epidemiology. Stochastic growth rates and complex term structures are a given in that field. So our hope is that this isn't the last use of Fixed Income modeling techniques for disease modeling.

\include{tractable}

\include{physical}

\bibliography{pandemic.bib}
\bibliographystyle{plain}

\end{document}

%% file: overview.tex
\section{Introduction}

We take an pragmatic approach to modeling epidemics subject to variation and random evolution. We show that modeled quantities can be computed as corrections to those that would be computed in deterministic homogeneous counterparts. We appeal to de Finetti's Theorem to justify the surprising generality of this strategy, despite the fact that the models we consider are simple mixtures of models - each one of which might not, in isolation, be considered plausible at the fine scale of neighbourhoods.  

\subsection{A meta-population model}

Our ansatz is a collection of non-interacting regions, termed cities for the sake of exposition. Within each city we imagine there are strongly interacting sub-populations called neighborhoods.\footnote{Lack of interaction between cities is an acknowledged limitation of the setup. But on the other hand different interpretations of populations and sub-populations may be employed - such as smaller regions or even cohorts of people.} 

We assume each city shares the same stochastic generative model. The path taken by each city's epidemic will be different, but we assume that mean quantities for a stochastic model for a city can be viewed as equivalent to an aggregate model for many cities - as if all paths are played out at once. 

\subsection{Measurements}

We consider model observables, or functionals, that are expectations with respect to the model for one city. We focus in particular on two mean quantities:
\begin{itemize}
    \item The overall (mean) early stage growth rate
    \item The mean across cities of the number of susceptible people at peak infection - where peak infection time is defined city by city.
\end{itemize}
 The second quantity is simpler to compute than a contemporaneous measure of peak infection, due to peak infection occurring at different times in different cities. 

\subsection{Approach via de Finetti's Theorem}

Our approach is to consider only functionals, such as those listed above, that are symmetric with respect to interchange of neighborhoods. We consider two models providing the same mean observable quantities to be equivalent, and to belong to the same equivalence class (orbit). For example, here are three equivalent expressions for the expected number of infections after three days. 
$$
       \overbrace{ E_{\phi^*}\left[ i(t=3) \right] }^{nature's\ true\ model} 
       = \overbrace{  E_{\psi=U \phi^*}\left[ i(t=3) \right] }^{nature's\ symmetrized\ model} 
       = \overbrace{  E_{\phi \in O(\phi^*) }\left[ i(t=3) \right] }^{any\ model\ in\ same\ orbit  } 
$$
We will describe this chain of equalities with slightly more formality. For now, it suggests that we may be well served by characterizing the orbit generated by a symmetry group which passes through nature's true model - perhaps by finding a representative of each orbit. 

To that end, we attempt to span the space of equivalence classes by considering only exchangeable models (there is precisely one on each orbit). Then, within that class, we only consider models that are mixtures of even simpler models. The simpler models assume neighborhoods are independent and identically distributed (iid). It may be clear to the statistical reader that this approach is inspired by de Finetti's Theorem and its variants for finite collections of variables. 

\subsection{Convexity adjustments}

For concreteness, and risking the ire of de Finetti, we shall further narrow things down by choosing a specific type of iid model. Our choice is a stochastic variant of the SIR compartmental model in which infections rates follow an Ornstein-Uhlenbeck process. This choice is well motivated both by empirical epidemiological literature and a physical model for repeat contacts. It also takes advantage of the connection to short rate models for interest rates, and we use the Vasicek bond pricing formula with negative interest rates to compute mean early stage growth.

In a similar fashion we can estimate mean peak infection using properties of the growth process. Moreover we can relate these quantities to their classic deterministic equivalents by multiplicative corrections that we call convexity adjustments. For example a crude but useful rule of thumb to account for stochastic evolution of infection rate is
$$
       susceptible\ multiplier = 1 + \frac{relative\ variation}{R}
$$
where the relative variation is the standard deviation of infection rate divided by mean infection rate. We advocate the use of convexity adjustments as a means of rapidly locating the orbit of nature's true sub-population model - by which we mean the orbit induced by permutations of neighborhoods. 

\subsection{Empirical variation}

This paper assumes variation in infection rate is plausible, but only indirectly adds to the evidence for this insofar as it may explain peak infection rates. It is widely believed that infection rate ($\beta$ as we shall refer to it) exhibits variation across various dimensions because it commingles a variety of factors driving the number of people who will be infected in unit time. Some are biological. Others are behavioral. 

Infection rate as defined in a compartmental model may be influenced by behavior and patterns of socializing that differ from place to place \cite{DelValle2007MixingNetworks}. At the individual level age \cite{Zhang2020ClinicalChina} dyspnea \cite{Wu2020RiskChina} cardiovascular disease \cite{Chen2020AnalysisCOVID-19}, pregnancy
\cite{Yu2020ClinicalStudy}, \cite{Zhang2020AnalysisProvince} and other factors may influence outcome and transmission likelihood. If airborne transmission \cite{Read2008DynamicDisease} is a key factor, or if supply of ultraviolet light \cite{Kowalski20202020Susceptibility} is important, these factors will play a bigger role in some parts of a city than others - even in one apartment versus another. Humidity \cite{Wang2020HighCOVID-19} is considered important. 

Variation in transmission rates may be driven by culture. A reasonable hypothesis is that household size and mixing of generations therein may be an important dynamic. Economic variation may be a driver also. Some may be better placed to afford protective measures, or voluntarily reduce hazardous work. 

A recent study of SARS CoV-2 in Iran suggests that wind speed may reduce infection in addition to other climatological factors \cite{Ahmadi2020InvestigationIran} such as temperature, which is said to drive down infection rates by three percent for every one degree increase \cite{Wu2020EffectsCountries}. Another intriguing possibility is that existing vaccinations may be associated with dramatically lower incidence \cite{Klinger2020SignificantlyAnalysis}. If true, those with the vaccinations and those without will constitute two very different groups with two very different turning points. Population density is another prime culprit because of its inherent plausibility and dramatic variation. We will devote additional time to population density in Section \ref{sec:shape}.

%% file: definetti.tex
\section{An approach inspired by de Finetti's Theorem}
\label{sec:definetti}

\begin{figure}
    \centering
    \includegraphics[scale=0.32]{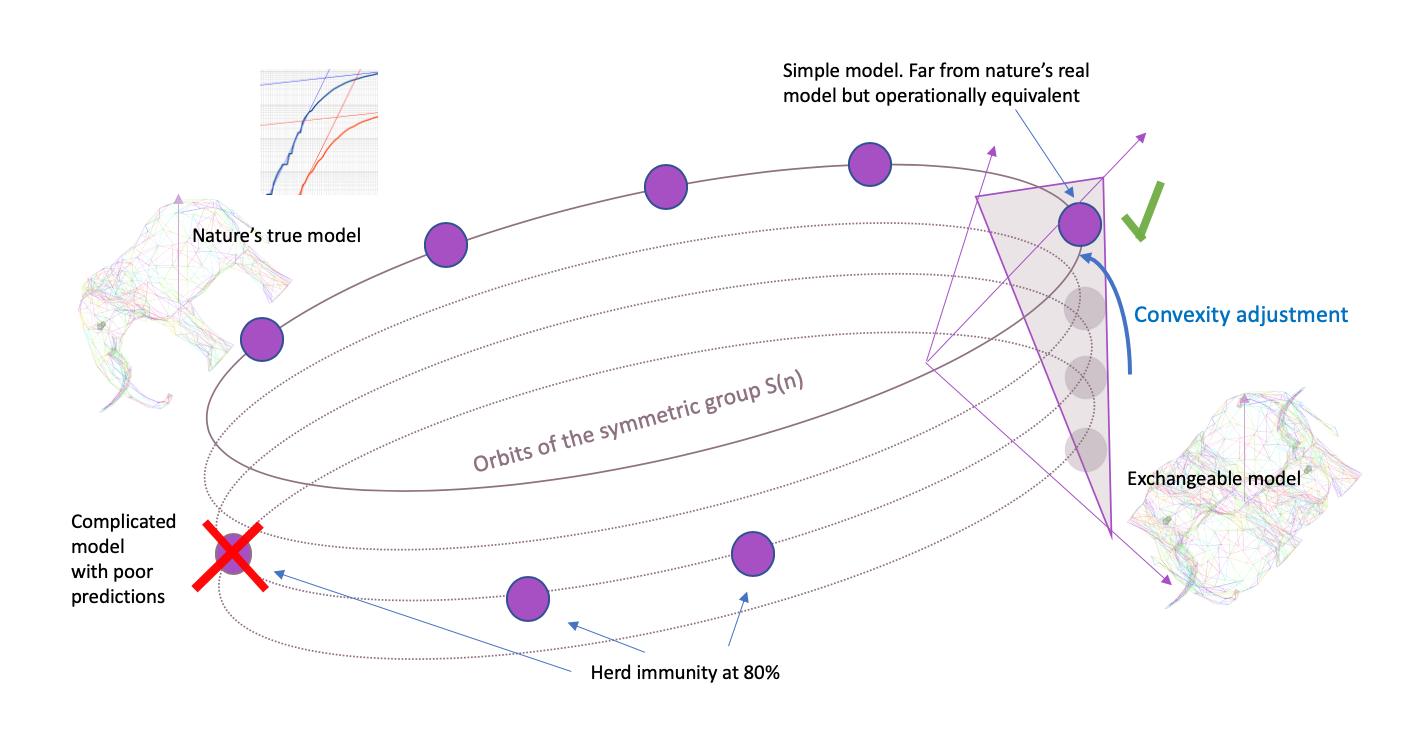}
    \caption{Outline of the strategy adopted in this paper. When only symmetric functionals of a model are required, we are motivated to consider representatives from orbits of the symmetric group only, rather than the full model space. Simple mixtures of models in which neighborhoods are independent might be used to quickly traverse a collection of exchangeable models, one of which might provide the same answers as nature's true model - a model we are unlikely to ever locate. Convexity adjustments relate stochastic model fucntionals to their deterministic counterparts, allowing for a potentially rapid resolution of modeling conundrums precipitated by early onset heard immunity.}
    \label{fig:orbits}
\end{figure}

Our strategy is represented stylistically in Figure \ref{fig:orbits} and is hopefully clear in broad brush terms from the introduction. Here we go into more detail to avoid ambiguity, though the reader familiar with exchangeability and finite versions of de Finetti's Theorem might safely skim parts of this Section. 

\subsection{Invariant functionals}

Consider the following questions that might be asked of the model
\begin{enumerate}
    \item When will peak infection occur?
    \item What is the distribution of case counts after seven weeks?
    \item What is the mean growth rate in the first three weeks?
    \item How many people will remain susceptible when infection peaks?
    \item How many cases will there be in the limit $t\rightarrow \infty$?
\end{enumerate}
Next imagine a city (more generally some population) is divided up in some manner (we use the term neighbourhoods). We imagine some partition $P_1,\dots, P_n$ and some quantity of interest $X_1,\dots, X_n$ measured on each. We suppose, however, that only aggregates such as $X = X_1 + \dots X_n$ matter for decision making purposes. More generally any symmetric function can be allowed, such as the sum of the squares of infections in each neighborhood. 

\subsection{Symmetric group}

We formalize this notion with the symmetric group of permutations. Under consideration is a space of models (joint distributions) on $n$ variables. There is an operation on this collection of models induced by permutations $\pi$ in the symmetric group $S(n)$ acting on $S=\{1,\dots,n\}$. Namely if $\phi$ is any joint distribution over $X=(X_1,\dots,X_n)$ then $\pi \cdot \phi$ is the function that when applied to $x=(x_1,\dots,x_n)$ yields
\begin{eqnarray*}
    \left( \pi \cdot \phi \right) (x_1,\dots,x_n) & = & \phi \left(\pi(x)  \right) \\
                                   & = & \phi\left(x_{\pi(1)}, \dots x_{\pi(n)} \right)
\end{eqnarray*}
As is well appreciated these actions form a group. If $\pi'$ is another permutation then applying the definition twice
\begin{eqnarray*}
    \pi' \left (\left( \pi \cdot \phi \right)\right) (x_1,\dots,x_n) & = & \left( \pi \cdot  \phi \right) \left( \pi'(x)   \right) \\
     & = & \phi\left( \pi\left( \pi'(x) \right) \right)  \\
     & = & \phi\left( (\pi' \circ \pi )(x)\right)  \\
     & = & \left((\pi'\circ \pi )\phi\right)(x_1,\dots, x_n)
\end{eqnarray*}

\subsection{Exchangeable models as orbit representatives}

Let $U$ denote the symmetrizing operator 
$$
    U := \frac{1}{n}\sum_{\pi \in S(n)} \pi 
$$
that averages any function over all possible permutations.\footnote{The notation $U$ for the symmetrizing operator comes from $U$-statistics.} This operator remains unchanged when composed with any permutation since 
$$
     \sum_{\pi \in S(n)} \pi = \sum_{\pi \in S(n)} \pi \circ \pi'
$$
given that the right hand side is a reordering of the sum on the left. Under the action of $U$ any model $\phi$ is mapped to an exchangeable model $U(\phi)$ meaning that it is left unchanged by the action of all elements of the symmetric group. Furthermore if $\psi_1$ and $\psi_2$ are both exchangeable and
$\psi_1=\pi \psi_2$ for any
permutation $\pi$ then 
$$
  \psi_1 = U \psi_1 = U \pi \psi_2 = \sum_{\pi' \in S(n)} \pi' \pi \psi_2 = U \psi_2 = \psi_2
$$
showing that there is one and only one exchangeable model per orbit.\footnote{In the second to last equality we are using the fact that every element of the symmetric group can be considered a permutation of the group itself, in turn because all elements are invertible.} These orbits, such as 
$$
O(\phi) := \left\{ \pi(\phi) \right\}_{\pi \in S(n)}
$$
are for us a useful decomposition of the set of possible models, and the considerations above show that the set of exchangeable models traverse the collection of all orbits. 

\subsection{Proxy models}
\label{sec:proxy}

Next we formalize the notion that an important planning question for an epidemic might be invariant to permutations. Decision making is said to be driven by functionals of the model. Denoting the space of all models by $\mathcal{M}$, any functional
$$
      f : \mathcal{M} \rightarrow \mathcal{R}
$$
on models (joint distributions) can also be acted on by a permutation $\pi$.\footnote{We follow a convention of defining this action in a contragredient manner, but this is choice is not essential to the argument. The functionals need not be scalar valued but in practice we can take them on one at a time.} This action is
$$
   \pi(f) ( \phi ) :=  f\left( \pi^{-1}(\phi) \right)
$$
and we can define a symmetrizing operator, also denoted $U$, on this space. It may be suggestive to write the application of the functional in bra-ket notation:
$$
     < \pi f, \phi > = < f, \pi^{-1} \phi >   
$$
When it is the case that $f$ is a symmetric functional, meaning $U(f)=f$, we have
\begin{equation}
\label{eqn:apply}
     < f, \phi > = < U f, \phi > = < f, U \phi >
\end{equation}
where we appeal to the fact that the sum comprising $U$ is left unchanged when we replace every permutation with its inverse, since this amounts to a mere reordering of terms. Equation \ref{eqn:apply} is the key to our suggested approach, for if we now suppose nature has a true model $\phi^*$ and we observe $f(\phi^*)$, we also know this value is shard by the same functional applied to a symmetrized version of nature's model. 
$$
      f( \phi^*) = f( U \phi^* )
$$
On the surface this is perhaps a controversial way to solve the challenge of how to create an accurate model that is close to reality (i.e. by not really trying). But seemingly what matters is not proximity to truth but proximity to the orbit of the truth. A limitation is that this approach leaves partly unaddressed the challenge of assessing the reasonableness of a {\em choice of orbit} (beyond its ability to match empirical data) a task not necessarily made obvious by any given representative of that orbit. 

\subsection{A pseudo-basis for exchangeable models}

However, we know that the exchangeable models constitute a traversal set. This particular traversal set may bring an additional advantage provided by, or at least inspired by, de Finetti's Theorem. 

In Section \ref{sec:vasicek} we will consider a collection of stochastic infection rate models. However that is not a requirement and the point can be made by considering the more familiar constant parameter SIR model \cite{1927AEpidemics}. Suppose we allow $\beta$ to vary with each realization and leave $\gamma$ fixed for simplicity.\footnote{If using deterministic evolution seems out of spirit with de Finetti's Theorem, the reader may decide to add some simple perturbation of this model, for instance through the addition of a measurement error distribution on instrumented quantities $i(t)$, $s(t)$ and $r(t)$ - or perhaps a more elaborate scheme involving delays in measurement.} 

To avoid confusion between average quantities and varying we shall separate out the variation as:
$$
       \beta(\omega) = \bar{\beta} \eta(\omega )
$$
where $\bar{\beta}$ is a constant rate of infection and $\eta(\omega)$ is a function defined on the space $\Omega$ over which variation takes place. Assume the population density is a measure on the same space and we have both 
$$
     E_{\rho}[\eta] = \int_{\Omega} \eta d\rho =  \int_{\Omega} \eta(\omega) \rho(\omega) d \omega =1 
$$
and
$$
       \rho(\Omega) = \int_{\Omega} \rho(\omega) d \omega = 1 
$$
where $\rho(\omega)$ is population density.\footnote{The space need not be continuous. For example $\Omega_1$ and $\Omega_2$ may represent disjoint sets populated by $30$ percent and $70$ percent of the population respectively. There is no loss of generality since $\bar{\beta}$ can be rescaled.} Then $\bar{\beta}$ can be interpreted as a population mean infection rate and $\eta(\omega)$ is the variation of the same.  

If this model is applied to a neighborhood $P_k$ and at the same time applied to all other neighbourhoods independently, the city model will be exchangeable. In fact the city model will be iid with respect to neighborhoods. Let us denote such a model as
$$
     m\left( \cdot; \omega \right) = \Pi_{k=1..n} m_k\left(  \cdot; \omega \right)
$$
The mixture model satisfies
\begin{eqnarray}
\label{eqn:mixture1}
    \dt i(t) & = & -\eta(\omega)\bb i(t) s(t) \nonumber \\
    \dt i(t) & = & \eta(\omega) \bb i(t) s(t) - \gamma i(t) \nonumber \\
    \dt r(t) & = & \eta(\omega)\gamma i(t) 
\end{eqnarray}
once $\omega$ is chosen. Assume now that $\eta$ takes on the law of a known distribution - preferably one with closed form moment generating function or harmonic mean for reasons that will become clear. We allow $\theta$ to summarize parameters controlling $\eta$. For example $\eta$ might be a rescaled Beta distribution and $\theta=(\alpha,\beta)$, overloading the use of $\beta$ just for a moment. Denote a mixing of models induced by $\eta$'s law as    
\begin{equation}
\label{eqn:span}
    m_{\theta} = \int m(\cdot, \omega) d\rho 
\end{equation}
This model is a mixture of independent, identically distributed models. The integration over $\eta$ will induce dependence between the neighborhoods. 

The question arises as to whether there is a $\theta$ for which $m_{\theta}$ can serve as a proxy for the truth. As per our discussion, this will be true if nature's model lies in the same orbit and, since $m_{\theta}$ is exchangeable, this also implies that 
$$
   m_{\theta} = U( \phi^* )
$$
for nature's true model $\phi^*$. 

We provide no guarantee that every exchangeable model can be represented in this fashion (unless $\rho$ is allowed to be a signed measure, which some readers may wish to consider) but as $\theta$ varies we may nonetheless traverse a reasonably representative collection of orbits. If the orbit of nature's true model exhibits strong positive correlation between neighborhoods - which in the case of an epidemic it almost certainly does - our chances that it exists on the orbit $O(m_{\theta})$ for some choice of $\theta$ go up substantially.

%% file: convexity.tex
\section{Convexity adjustments}
\label{sec:convexity}

We now come to the chief advantage of using mixtures of simple models. Predictions of the mixture model can often be written as multiples of those for deterministic models. The multiplier captures the impact of Jensen's Inequality, and can often be expressed in closed form in which case it is particularly easy to invert. Our approach makes it difficult to create very badly mispecified models. 

As an aside, Jensen's Inequality has cropped up in numerous biological contexts. The effect of daily fluctuations in temperature on Rhodnius prolixus, a vector of Chagas disease, was studied by Roalandi and Schilman \cite{Rolandi2018TheProlixus}. A Jensen's Inequality effect was noted in regard to prevalence of sigma virus and host Drosophila melanogaster in Georgia peach stands \cite{Wayne2011TheMelanogaster}. Closer to our interest, the nature of bias introduced in cohort-based models for disease progression was considered by Elbasha and Chhatwal \cite{Elbasha2015CharacterizingModels}. 

\subsection{Peak infection}

We continue with the same notation as in Section \ref{sec:definetti}.
Suppose in a given realization $\omega$ peak infection occurs at time $t^*$. At this moment the proportion of the population yet susceptible is $s(t^*)$. In the classic SIR model we expect 
$$
        s(t^*) = \gamma/\beta 
$$
but with a mixture model this quantity gets scaled by a quantity we know exceeds unity (by Jensen's Inequality).\footnote{Fix two people Irene and Sam, respectively infectious and susceptible. Let $\beta$ denote the instantaneous probability (per unit time) that Irena both interacts with, and also infects, Sam. Let $\gamma$ denote the instantaneous probability that Irene ceases to be contagious. In more general compartmental models $\gamma$ may comprise a sum of hazard rates for recovery, death, secure isolation or other states.  In the absence of any other dynamics influencing the number of susceptible people\footnote{The tally of people may be indicative of the severity of an epidemic, even if the peaks are not reached at the same time.} we have the accounting identity
$$
        \beta s(t^*) = \gamma
$$
and an expression for the vulnerable fraction of society
\begin{equation}
\label{eqn:herd}
       s(t^*) = \frac{ \gamma} {\beta }
\end{equation}
as a ratio of instantaneous rates.} We have instead

$$
       s(t^*) \approx  E\left[\gamma/\beta \right] =  \gamma/\bb E\left[\frac{1}{\eta}\right]
$$
The ratio of the mixed versus unmixed model with $\beta=\bb$ is a multiplicative  convexity adjustment, 
$$
       J = E\left[\frac{1}{\eta}\right]
$$
where we would be tempted to use the notation $H$ for herd immunity were it not for the confusion this would cause ($J \approx H^{-1}$ not $H$, where $H$ is standard notation for the harmonic mean). Furthermore this is only approximate.\footnote{A slightly more careful analysis would consider the case when $\beta(\omega)<\gamma$ and the first order condition is no longer relevant. A better approximation is provided by 
$$
    s(t^*) = \gamma/\beta E\left[\frac{1}{\max(\eta,\eta_{min})}
            \right]
$$
where $\eta_{\min}=\gamma/\beta$. For smaller variation this may be splitting hairs. When one considers that many prominent epidemic models failed to predict peak infection to within a multiple of two, it seems reasonable to surmise that in practice, any halfway-reasonable convexity adjustment is likely to move the modeler significantly closer to a well specified model. We use the simpler harmonic mean approximation here because it is so easy to invert, thus giving the researcher a simple ballpark estimate of variation.}  

To illustrate, suppose that the law of the variation function $\eta$ is a scaling of a Beta distributed random variable.\footnote{We need to scale so that the mean of $\eta$ is unity. Let $\Omega=(0,\beta/\alpha)$ and $\rho=\frac{1}{\beta/\alpha}$ be the uniform measure and $\nu$ be defined by the inverse cumulative distribution of a beta distributed variable that has been scaled by $\beta/\alpha>0$.}  The mean of $Beta(\alpha,\beta)$ is $\frac{\alpha}{\alpha+\beta}$ and the harmonic mean exists for $\alpha>1$ and $\beta>0$ which we assume. The harmonic mean is $\frac{\alpha-1}{\alpha+\beta-1}$ and the ratio of the mean to harmonic mean is
$$
   J = \frac{\alpha}{\alpha+\beta} \frac{\alpha+\beta-1}{\alpha-1} =\frac{\alpha^2+\alpha\beta-\alpha}{\alpha^2+\alpha\beta-\alpha-\beta}
$$
which provides the convexity adjustment in closed form. 

\subsubsection{Unormalized variation functions}
\label{sec:unnormalized}

A brief pragmatic comment on calculation of convexity adjustments. Suppose $\eta = c \varphi$ for some constant $c$ and unormalized variation function $\varphi$. Then    
$$
   J = \frac{1}{c} \int_{\Omega} \frac{1}{\varphi} d\rho =  \frac{  \int_{\Omega} \frac{1}{\varphi} d\rho }{ \int_{\Omega} \varphi d\rho  } 
$$
and it is often more convenient to compute a ratio of two unnormalized integrals than worry about constants. In particular we will consider cases where $\varphi \propto \rho^{\alpha}$ and thus
$$
      J(\alpha) = \frac{ \int_{\Omega} \rho(\omega)^{1-\alpha} d\rho } 
                       { \int_{\Omega} \rho(\omega)^{1+\alpha} d\rho                              }
                       = \frac{I(-\alpha)}{I(\alpha)}
$$
for some integral $I$ with parameter $\alpha$. Evidently it is not necessary to carry through all constants in the integrals in order to determine $J(\alpha)$ or a ratio of two $J(\alpha)$.

\subsubsection{Interpreting reported R values based on peak infection}

With these thoughts in mind, if a reported number ($\hat{R}$ say) is known to be derived from a peak infection estimate, which is to say
$$
  \hat{R} = 1/{\hat{s}(t^*)}
$$
for some estimate $\hat{s}(t^*)$ of peak $s$ we should be careful to interpret this directly as an estimate of the population size and not a link to mean infection rate via the SIR identity $R_0=\gamma/\beta$. Even if we bravely identify $\sT$ with $s(t^*)$ we would be better served by 
$$
    \bar{\beta} = J \gamma \hat{R} 
$$
instead of $\beta=\gamma R_0$. The mean rate of contagion $\bar{\beta}$ we infer is {\em higher} than a constant $\beta$ that might be inferred using a homogeneous model. 

\subsubsection{Peak infection at different times}

In the preceding discussion of peak infection we implicitly assumed critical times $\mathcal{T^*} = \{ t_{\omega} \}_{\omega \in \Omega}$ in the evolution of an ensemble of classic SIR compartmental models. Really, peak infection globally occurs when 
$$
   i'(t_{\omega}) = 0 =  \int_{\Omega}  \bar{\beta}\eta(\omega)  i(t_{\omega},\omega)s(t_{\omega},\omega) \rho(\omega)  - \gamma i(t_{\omega},\omega) \rho(\omega) d\omega 
$$
and it is bold to approximate this by assuming balance holds for each $\omega$ simultaneously:
$$
            s() = \frac{\gamma}{\bar{\beta}\eta()}
$$
Here $s(\omega)\rho(\omega) d\omega = s d\rho $ is interpreted as the count of susceptible people in a sub-population at this turning point $t_{\omega}$. This gives rise to the approximation 
$$
  s(t^*) \approx \sT = \int s(t_{\omega}) d\rho = \frac{\gamma}{\bar{\beta}} \int_{\Omega} \frac{1}{\eta}  d\rho 
$$
However the first approximation may be poor. In the appendix we consider a further convexity adjustment taking into account the differing times of peak infection. 

On the other hand as variation is increased the set $\mathcal{T}$, the times $t_{\omega}$ at which individual peaks are reached, will become more disperse so the estimate $\hat{s}(t^*)$ will be lower than it otherwise would be with no variation. In fact we will have
$$
        \hat{s}(t^*) =  \sT/F = \frac{1}{F} \int s(t_{\omega}) d\rho  
$$
for some ratio $F>1$ depending on choice of $\bb$ and $\eta$ (also $\gamma$ and $i(0)$).\footnote{The mnemonic here is that $F$ stands for Fourier. For readability all ratios $F$, $J$ and $G$ to follow are defined so as to be greater than $1$.} Thus a more careful inference would be 
$$
     \hat{R} = \frac{1}{\hat{s}(t^*)} = \frac{1}{F \sT} = \frac{\beta_h}{
     F \gamma} =  \frac{1}{F} \frac{1}{J} \frac{\bb}{\gamma}
$$
leading to 
$$
    \bb = F J \gamma \hat{R}
$$
in place of the usual $\beta=\gamma R_0$ in the SIR model.

\subsection{Early stage growth}

Convexity adjustments are also available for early stage growth. We consider mean growth starting with the same proportion of initially infected people $i(0)$ where again infection is drawn at random according to $\beta(\omega) = \bar{\beta}\eta(\omega)$ as before.\footnote{We are attempting to determine properties of an exchangeable model that lies on the same orbit as a realistic one. For a discussion of interacting groups and early stage growth see \cite{Nishiura2010Pros2009} for example.} We further assume $s(t)\approx 1$ since we assume the measurement takes place in the early stage of the epidemic. So approximately
we have
$$
  i'(t;\omega) = \bar{\beta}\eta(\omega) i(t) - \gamma i(t)
$$
Thus by integration
$$
    \frac{i(t)}{i(0)}  =  \int \frac{i(t;\omega)}{i(0)} d\rho(\omega) \\
          =    e^{-\gamma t}  E_{\rho} \left[ e^{t \bar{\beta}\eta} \right]    
$$
We might refer to the quantity 
$$
        \beta_g := \frac{1}{t} log E_{\rho} \left[ e^{t \bar{\beta}\eta} \right] 
$$
defined for some time $t$ that is well before herd immunity, as the {\em growth effective infection rate}. It plays a similar role to $\beta$ in the constant parameter SIR model, at least as far as growth in aggregate infections is concerned. The growth effective infection rate is a multiple of $\bb$, 
$$
          \beta_g = G(t) \bb 
$$
where we introduce $G$ and note that again by Jensen's Inequality $G(t)>1$. The growth ratio $G$ will depend on the parameters of $\eta$ and the time $t$ marking the end of the interval $(0,t)$ where we measure growth. 

It will be apparent that $G$ is closely related to the moment generating function of the variation function $\eta$. In financial terminology it also plays a role analogous to a yield. We will draw a closer connection to finance in Section \ref{sec:vasicek}. Note for now that dividing by $\bb$ we have
$$
    G(t) = \frac{ \log E_{\rho}[ e^{t\bb \eta}]}{\bb t}
$$
showing that adjusted growth as a multiple of $\bb$, when time is measured in units of $1/\bb$, is equal to the extent to which the logarithm of the moment generating function of $\eta$ grows faster than linear time. In those more sensible time units
$$
       G'(t') = \frac{ \log E_{\rho}[ e^{t'\eta}]}{t'} 
$$
is slightly more convenient expression of $G$ with $t'=\bb t$. 

\subsubsection{Gaussian approximation}

For gaussian $\eta$ 
$$
      E_{\rho} \left[ e^{t \bar{\beta}\eta} \right]    \approx   e^{t \bb \overline{\eta} +\frac{t^2}{2} \bb^2 E[(\eta-1)^2] }  
$$
is exact and this approximation may be useful elsewhere. The exponent is the variance $\sigma^2$ of the variation function $\eta$. Thus measured growth 
to time $t$ will be 
$$
   g(t) = \bb -\gamma + \frac{t}{2}\bar{\beta}^2 \sigma^2 
$$
and effective $\beta_g$ given by 
$$
       \beta_g = \bb \overbrace{\left( 1+\frac{t}{2}\sigma^2 \right)}^{G(t)}
$$
although this demands care in interpretation for non-gaussian variation. The issue is that $\bb t$ may not be small so approximations for small times may have limited utility. Early stage growth is often measured after a few multiples of $1/\bar{\beta}$ to minimize small sample effects, censoring and related measurement inaccuracies. It may also be prudent to choose $t$ larger than $1/\gamma$, the typical recovery time, to reduce the impact of short time effects such as an initial, rapid decrease in novel interactions. 

If $\eta$ varies by a large percentage over some regions of the probability space we must be additionally careful. Therefore in what follows we assume $t$ is several multiples of $1/\bar{\beta}$. To the extent that analytic approximations are useful one might steer towards the expansions around $t\bar{\beta}=\infty$ rather than $t=0$. We shall also compute $G$ for several distributions of $\eta$ in Section \ref{sec:tractablegrowth}. 

\subsubsection{Growth adjustments under stochastic evolution}
\label{sec:fixedincome}

Thus far we have considered deterministic evolution. However to prepare for Section \ref{sec:vasicek} we briefly remark that this is not a prerequisite and we can allow stochastic dynamics. In the field of quantitative finance, considerable work has gone into the finding of solutions of the form
$$
   u(t,x_0) = E\left[ \exp\left(\int_0^t x_s ds \right) | x(0)=x_0 \right]
$$
where $x_s$ is a stochastic process and the integral can be viewed as a Feynman-Kac formula solving a diffusion equation. One interpretation arises when $x_s$ is a short rate (instantaneous rate of interest) and the integral is the balance in an account after time $t$ when all interest is reinvested. Assuming a convenient expression or method of calculation, the growth convexity adjustment is facilitated by the fact that 
\begin{equation}
    \label{eqn:convexity}
    \frac{ E\left[ \exp\left(\int_0^t x_s ds \right) | x(0)=x_0 \right]  }{
       \exp\left(E\left[ \int_0^t x_s \right] ds \right) }
       = \frac{ E\left[ \exp\left(\int_0^t x_s ds \right) | x(0)=x_0 \right]  }{
       \exp\left(\int_0^t E[x_s]  ds \right) }
\end{equation}
is tractable (where the equality glosses over some technicalities). 

More frequently, at least in the fixed income literature, formulas as provided for the closely related quantity 
$$
   u(t,x_0) = E\left[ \exp\left(\int_0^t -x_s ds \right) \right]
$$
whose interpretation is the price of a bond with maturity $t$. The insurance and failure rate literature also provided examples of convenient choices of stochastic process $x_s$ where $x_s$ is interpreted as an instantaneous rate of failure of a machine, or default of a bond issuer and the integral represents a probability of survival (either a real world probability or more likely, in the case of credit default swaps, a pricing measure). 

For some solutions there is a trivial connection between the moment generating function and the survival function. We give an example in Section \ref{sec:vasicek} of using $-x$ in place of $x$ to compute mean growth for an epidemic.\footnote{Depending on the model in question, reflection under the x-axis may not be possible. However it may be possible to recycle existing analytics in other ways such as reflection in time.}

\subsubsection{Limitations}

The convexity adjustment $G$ ratios assume $s(t)\approx 1$ and will be invalid if the susceptible population rises to an appreciable level for any appreciable mass of $\Omega$. 

\subsection{Combining convexity adjustments longitudinally}

Having considered the impact on both early stage growth and peak infection, we turn to the questions of how observations of the former might inform the latter - and how one might find models which are capable of fitting both types of evidence for some choice of parameters. 

\subsubsection{Forecasting peak infection}

The mean infection rate $\bar{\beta}$ and variation $\eta$ describe the system and thus must relate the two, but they are not directly observed. Through a homogeneous SIR model lens we might
\begin{enumerate}
    \item Estimate $R_0$ by some means
    \item Forecast $s(t^*) = 1/R_0$. 
\end{enumerate}
And we note that theoretically $R_0=\gamma/\beta$. The evidence from Sweden seems to suggest there is something askew with this logic. 

Here is a possible explanation. Assuming $\gamma$ is known and the true system behaves is as we have described it - as a mixture with varying $\beta$ - the first step will in fact produce an estimate of the growth effective infection rate $\beta_g$ due to the relation $R_0 = \gamma/\beta_g$. Here the denominator is related to $\bb$ and $\eta$ by 
$$
      \beta_g = G(t) \bb
$$
as we have seen. We have also determined that the susceptible population at peak $s(t^*)$ relates to $\bb$ and $\eta$ via 
$$
      s(t^*) = \beta_h / \gamma = \frac{1}{J}\frac{\bb}{\gamma}
$$
Thus we might write  
\begin{equation}
\label{eqn:peakinfection}
    s(t^*) = \frac{1}{J} \frac{1}{G} \frac{\beta_g}{\gamma} =  
    \frac{1}{J} \frac{1}{G} \frac{1}{R_0}
\end{equation}
The threshold number of infections will be higher due to two applications of Jensen's Inequality. 

One can add a third if we also consider the impact of peak infection at different times and are able to estimate this
$$
   s(t^*) =  
   \frac{1}{F} \frac{1}{J} \frac{1}{G} \frac{1}{R_0}
$$
We emphasize that the term $G(t)$ will be a function of the interval $(0,t)$ over which a measurement of early stage growth is made.

\subsubsection{Avoiding egregious model misspecification}

Assume now some empirical data at different stages of epidemics and a desire to build models that can, for some choice of parameters, explain multiple types of observation coherently. In particular suppose we are presented with data for the two functionals we have paid attention to: early stage growth and peak infection susceptible population. 

We do not feel qualified to accurately quantify mispecification given the miriad issues involved with COVID-19. But let us suppose, for the sake of discussion, that when a constant parameter SIR model is used to infer $R$ at peak infection the ratio of $\beta$ to $\gamma$ is approximately $5/4$. On the other let us also suppose, again just for illustration, that this ratio implies $1.7$ times less early stage growth than is implied by reproduction numbers on the lower end of the spectrum.\footnote{In the case of constant $\gamma$ the time until an infectious person transmits to another is exponentially distributed. This leads to a relationship between reproduction number $R$, growth $r$ and recovery $\gamma$
$$
      R = 1 + \frac{r}{\gamma}
$$
consistent with $R=\beta/\gamma$ and 
$$
     r = \beta-\gamma
$$
as we expect. When variation is introduced into $\gamma$ the reproduction number will be reduced relative to the reproduction number computed using the mean of $\gamma$. However we shall limit our attention to convexity adjustments for the variable $\beta$, referring the reader to Wallinga and Lipsitch for further discussion of mean generation times \cite{Wallinga2007HowNumbers}.} 

We propose to use the relationships established in Section \ref{sec:convexity} to quickly identify the parameter $\theta$ which controls the variation in $\eta$ and thus $\beta$. In other words, we solve
$$
   \overbrace{1.7}^{Swedish\ puzzle\ factor} =  \overbrace{\frac{mean\ growth}{SIR\ growth}}^{growth\ convexity } \times 
        \overbrace{\frac{arithmetic\ mean\ \beta(t^*)}{harmonic\ mean\ \beta(t^*)}}^{herd\ convexity}
$$
for $\theta$ or more abstractly, solve for $\eta$. We emphasize again that the variation implied by this procedure must be viewed in the light of de Finetti's Theorem. The mixing serves as a proxy for {\em dependence} between neighborhoods that are otherwise independent, so it need not represent a seemingly realistic level of actual spatial variation. 

Referring back to our original Figure \ref{fig:orbits}, it is apparent that inverting convexity adjustments can move us much closer to the correct orbit. This approach stands in stark contrast to other modeling strategies which involve modifications to elaborate simulations that attempt to model reality at a granular level (using geospatial demographic data, transport networks and so forth). These models may in some respect be closer to the truth. However they may easily be further from the true orbit. 

\subsubsection{An iterative procedure}

Inverting the convexity adjustments is unlikely to present serious difficulty but for concreteness, one simple iterative approach would simply assume equal convexity adjustments arising. Let's say the functionals $f_1$ and $f_2$ represent measurements and $J_1$ and $J_2$ the respective convexity adjustments, once might suppose $\gamma$ is known and $\bb$ and $\theta$ are to be determined. We assume a missing factor of $\lambda$ where, in the case of Sweden's puzzling herd immunity, we might set $\lambda=1.7$ say. As a first pass we express this model mispecification as
$$
      f_1( \phi(\bb_0,\theta ) ) \approx J_1(\theta,\bb) f_1( \phi(\bb_0,\eta\equiv 1 ) )   \approx \sqrt{\lambda} \overbrace{f_1( \phi^* )}^{observed} 
$$
where $\phi^*$ is nature's unseen true model but $f_1(\phi^*)$ is observed, and 
$$
        f_2( \phi(\bb_0,\theta ) ) \approx J_2(\theta,\bb) f_2( \phi(\bb_0,\eta\equiv 1 ) )   \approx \frac{1}{\sqrt{\lambda}} \overbrace{f_2( \phi^* )}^{observed}  
$$
and $\bb_0$ is some first guess of infection rate parameter taking into account the conflicting evidence.

For instance if threshold infection implies $\beta=1.25\gamma$ and growth implies $\beta=1.65 \gamma$ then we take $\beta^*=1.45\gamma$. Here $\phi^*$ is nature's true model and the right sides of these equations are observed quantities.\footnote{
This may well be the case at time of writing. If $f_1$ and $f_2$ represent measurements of early stage growth of an epidemic and threshold infection, as we have been discussing, then evidence from Sweden may be interpreted as an onset of herd immunity with $s(t) \approx 0.8$ corresponding to a ratio of $\beta$ to $\gamma$ in the SIR model of $5/4$, much lower than $R$ values implied by spread of the disease.} Each of these equations might be solved for $\theta$ independently, leading to a next guess at $\theta$ (say the average of the two) and a next value of $\lambda$. This simple approach is motivated by the observation that $J$'s may be independent of $\bb$ altogether or only weakly depend on $\bb$.

\subsection{Cross-sectional use of convexity adjustments}

Before leaving the topic of convexity adjustments we point our that longitudinal prediction is not the only application. Since variation driven adjustments provide corrections for measurable quantities relative to deterministic compartmental models, they can also provide a very convenient way to relate two different epidemics {\em at the same stage} subject to different conditions. 

Moreover, the convexity adjustments per se may not vary strongly with underlying assumptions used for the deterministic models. So it is possible that this sort of comparison may sidestep problems associated with calibration of compartmental models. To illustrate we consider threshold infection levels and once again the measurable of interest is the number of susceptible people when balance holds. 

\begin{figure}
    \centering
    \includegraphics[scale=0.6]{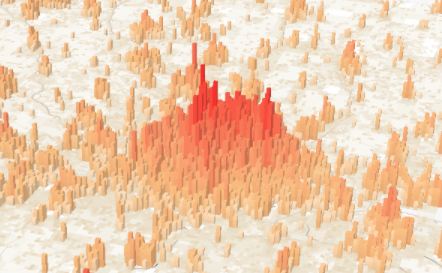}
    \caption{Population density for London.}
    \label{fig:london}
\end{figure}

\begin{figure}
    \centering
    \includegraphics[scale=0.4]{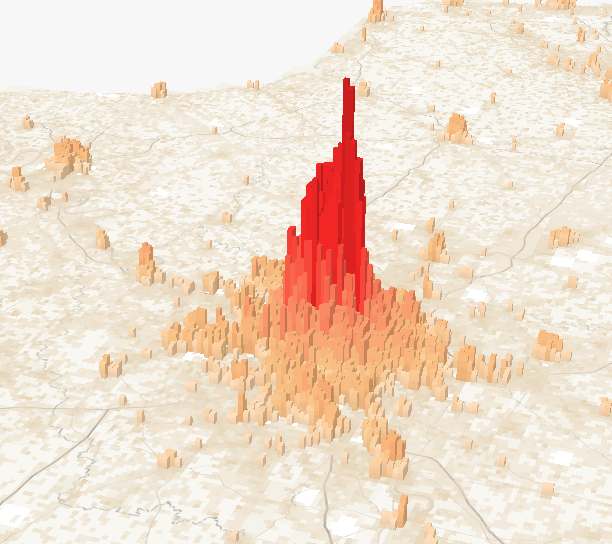}
    \caption{Population density for Paris}
    \label{fig:paris}
\end{figure}

To illustrate we assume variation $\eta$ is driven by population density. Though this is no the only source of variation it is certainly significant and moreover the degree of variation also varies considerably. This is manifest in the way population density takes very different {\em shapes}, referring to the geometry of population density. Figure \ref{fig:london} shows the population density for London, whereas the equivalent for Paris is depicted in Figure \ref{fig:paris}.\footnote{Population graphics provided by Matt Daniels of The Pudding.}  

Empirical study of population density and its effect on epidemic growth is a muddied picture with some authors appealing to the likelihood of a large impact (e.g. \cite{Tarwater2001EffectsDisease}) whereas in a study of pine beetles a thinning of the population did not noticeably change an epidemic's equilibrium  \cite{MacQuarrie2011Density-dependentStands}. Population fluctuations appear to drive pathogen spread in Niger \cite{Ferrari2010Rural-urbanNiger}. Different rates of microparasite transmission were observed in cities versus countries, providing indirect evidence of the role of density \cite{Grenfell1998CitiesMetapopulation}. HIV transmission reconstructed with phylogenetic methods revealed density dependence \cite{Leventhal2014UsingTransmission}. 

COVID-19 cases in China and the U.S. have seemingly concentrated in areas of high population density - though there are of course confounding factors such as proximity to international airports and movements of people. Without seeking to evaluate the evidence we suppose merely in this section that population density plays a role - possibly a weak one.\footnote{In fact the convexity adjustments work when infection is proportional to a power of population density less than $1$ as well as greater than $1$, so one might even use them under the seemingly unlikely scenario that density reduces infections.} 

Focusing on cross-section comparison, we suppose two towns: Town A and Town B. We assume the epidemic has long since run its course in Town A. We are given the task of estimating herd immunity levels and reproduction numbers for Town B. This town shares many of the same properties as Town A including the same mean contagiousness - which is to say the likelihood of a contagious person infecting another. The only difference between the two towns we can identify is that Town A's population profile is shaped like a bell curve whereas Town B's falls away more like an exponential distribution. 

\subsubsection{Rules of thumb for peak infection from population shape}
\label{sec:shape}
\label{sec:towna}

\begin{table}[h!]
\begin{tabular}{|l|l|l|l|}
\hline 
    Town A &   & Town B & \\
    \hline 
     Shape    & Gaussian & Shape & Exponential   \\
     Reproduction & 5.0 &  Reproduction   & $R_b$ ? \\
     Herd     & $h_a=0.8$ &  Herd & $h_b$ ?    \\
     Vulnerable & $v_a=0.2$ & Vulnerable  & $v_b$ ? \\
   \hline 
\end{tabular}
\end{table}

\begin{table}[h!]
\label{tab:coef}
\begin{tabular}{|l|l|l|l|}
\hline 
    Shape & Coef  & Shape & Coef \\
    \hline 
     Flat     & 0 & Skirt ($1/r^4)$ & 4   \\
     Gaussian & 2 & Gamma ($r exp(-r)$ & 4   \\
     Exponential & 4 & Gamma ($r^2 exp(-r)$) & 3  \\
     Skirt ($1/r^3$) & 6 & & \\
     \hline 
\end{tabular}
\end{table}
Based on {\em ratios of convexity adjustments} to be provided, we suggest the following way to answer this challenge. A calculations such as the following can be used to adjust for population shape:
$$
     v_b = v_a \left\{ 1+ \underbrace{0.1}_{power\ law}(\overbrace{4}^{shape\ b\ coef}- \overbrace{2}^{shape\ a\ coef} ) \right\} = 1.2 v_a = 0.24 
$$
from whence it follows that 
$
     h_b = 1- v_b = 0.76
$
and 
$
    R_b = 1/0.76 = 4.16.
$
This heuristic is derived from first order approximations of two convexity adjustments performed back to back. The reader may choose to view them as reproduction number interpretations chained as follows:
$$
     R_a \rightarrow R_{hom} \rightarrow R_b
$$
where $R_{hom}$ represents a reproduction number that that would exist in the absence of any variation. We probably never get to see $R_{hom}$, however, because there isn't a country, city or neighbourhood in the world lacking variation. Instead, we first {\em infer} a true underlying level of contagiousness in Town A by subtracting a convexity effect, and then add back the convexity correction for Town B. 

The coefficients provided in Table \ref{tab:coef} are simply the first terms in a Taylor expansion of the product of convexity adjustments. We provide some formalization of these calculus exercises. 

\subsubsection{Rotation invariant population densities and power law}

We identify the probability space with the plane $\mathcal{R}$. The link between density and contagiousness is muddied \cite{Tarwater2001EffectsDisease} \cite{Leventhal2014UsingTransmission} but even a weak relationship may be important. We shall for simplicity assume that contagiousness $\beta$ is a function of population density via a power law. 
\begin{equation}
      \beta(\rho) \propto \rho^{\alpha} 
\end{equation}
for some $0<\alpha<1$ and population density $\rho$. Further, we shall assume that density is a radial function from the origin and thus so is $\beta(\cdot)=\beta(r)$ is as well. We can write   
$$
            \frac{\beta(r)}{\bar{\beta}} = \eta(r) \propto \rho(r)^{\alpha} 
$$
As per our comments in Section \ref{sec:unnormalized} it is marginally cleaner to work with integrals of an unnormalized variation $\phi$ and determine the ratio of the integral. 

\subsubsection{Gaussian population density}

If a city's population density varies as $\rho(r,\theta)=\frac{1}{2\pi} e^{-r^2/2}$ as we move out a distance $r$ from the center then we take $\phi(r)=e^{-r^2 \alpha/2}$
$$
      \iint \rho \phi   =  \frac{1}{2\pi}  \int_0^{\infty} e^{-\frac{1+\alpha}{2}r^2} r dr = \frac{1}{{1+\alpha}}
$$
Recycling the same integral with $\alpha \mapsto -\alpha$ and taking a ratio:
$$
   J(\alpha) = \frac{1+\alpha}{1-\alpha}
$$
Merely as an example, $J(a=1/4) = 5/3$ if we use a power law coefficient loosely informed by Measles outbreaks in the United States.

\subsubsection{Exponentially distributed populations}

Next consider a city with more sprawl. Set $\rho(r,\theta)= \frac{1}{2\pi} e^{-r}$ to create a city whose density falls off more abruptly at first, but with a much longer tail. With $\phi(r)=e^{-\alpha r}$ we have 
$$
   \iint \rho \phi   =   \int_0^{\infty} e^{-(1+\alpha)r} r dr  = \frac{1}{(1+\alpha)^2}
$$
yielding 
$$
   J(\alpha) = \left( \frac{1+\alpha}{1-\alpha} \right)^2
$$
which is the square of the Jensen's ratio for a Gaussian city. Comparing with the Gaussian case and holding everything else constant, even a ten percent relationship between log contagiousness and log density might imply a twenty percent correction in the number of people who escape the virus when we translate the data from a Gaussian city to an Exponential one.  

More examples of shape adjustments are given in Section \ref{sec:tractableshape}

%% file: vasicek.tex
\section{Vasicek meta-population mixture models}
\label{sec:vasicek}

\begin{figure}
    \centering
    \includegraphics[scale=0.5]{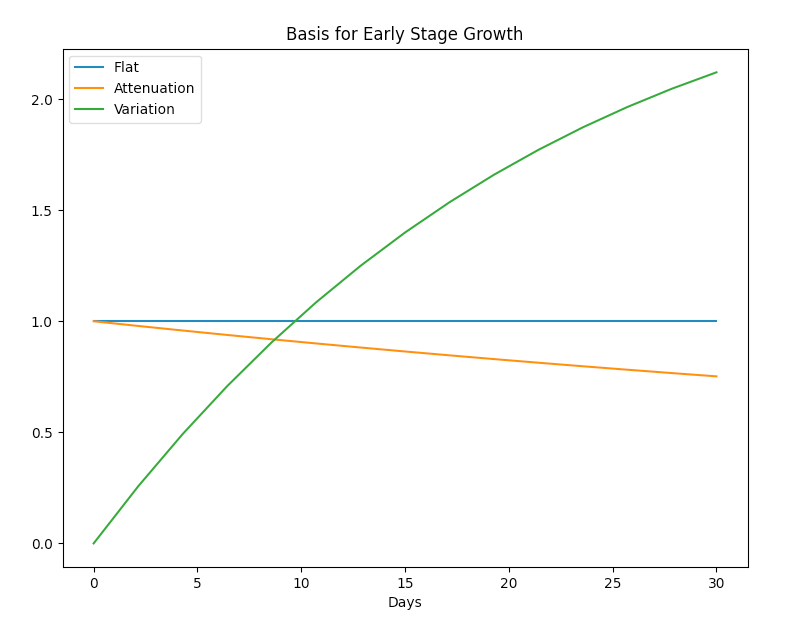}
    \caption{Caption}
    \label{fig:orbits}
\end{figure}

We have outlined a strategy using mixtures of 
independent identically distributed (iid) models that are carefully interpreted as representatives of equivalence classes of models as per Figure \ref{fig:orbits}. We now apply the approach using models where each neighborhood's epidemic follows a stochastic compartmental model. Infection rate follows an Ornstein-Uhlenbeck process, which implies early stage growth follows an Ornstein-Uhlenbeck process as well. 

This approach also adopts the suggestion in Section \ref{sec:fixedincome}, namely that early stage growth calculations can be borrowed from the fixed income literature. We call this a Vasicek compartmental model after the corresponding work on interest rates by Vasicek (in which short rate follow an Ornstein-Uhlenbeck process) and we borrow his formula for the price of a bond that pays no coupons. 

While this may be taken as a mere example and a somewhat arbitrary (suspiciously convenient) choice of basis, we provide in Section \ref{sec:justifyvasicek} some motivation for Ornstein-Uhlenbeck infection rates leaning both on empirical work \cite{Chowell2016MathematicalReview} and also a more theoretical justification in terms of a physical model for repeat contact probability \cite{Cotton2020RepeatDisease}. 

\subsection{Stochastic process for infection}
\label{sec:stochasticinfection}

In contrast to most of the discussion in Section \ref{sec:convexity}, we allow infection rate $\beta(t)$ to follow a stochastic process rather than a deterministic process. However in a manner similar to the preceding it will have parameters driven by $\omega \in \Omega$ once again, where $\Omega$ is an additional probability space having nothing to do with the evolution of the stochastic process driving $\beta$. A general example is
$$
     d\beta(t) = \kappa(\omega)\left(\beta_{\infty}(\omega) -\beta(t)\right) dt + \sigma_{\beta}^2(\omega) dW_t
$$
with initial condition $\beta(t=0;\omega)=\beta_0(\omega)$. Here $\beta_{\infty}$ is a mean reversion level and the rate $\kappa(\omega)$ controls how quickly reversion occurs. This is inserted into the SIR differential equations as before. Here the initial value of infection rate, variance, mean reversion level and pull $\kappa$ can all be varied with $\omega$. Notice that if we define 
$$
    r(t) = \beta(t)-\gamma
$$
then $dr(t)=d\beta(t)$ and $r(t)$ may be seen to satisfy a similar stochastic differential equation.  Add zero to the drift:
$$
   dr(t) = \kappa\left(\overbrace{\beta_{\infty}(\omega)-\gamma}^{r_{\infty}} - \overbrace{(\beta(t)-\gamma)}^{r(t)}\right) dt + \sigma_{\beta}^2(\omega) dW_t
$$
to see this, noting that $r(t)$ also has different initial condition $r_0=\beta_0-\gamma$.

\subsection{Iterated expectations}

We begin with an elementary observation that applies if evolution of $\beta(t)$ is unrelated to choice of $\omega$, or more accurately if it is conditionally independent. By the law of iterated expectations we have
$$
  E \left[ e^{\int_0^t r(s) ds} \right] = E_{\rho} \left[ \overbrace{E_{W}\left[ e^{\int_0^t r(s) ds} \right]}^{inner} \right] 
$$
and thus 
\begin{equation}
    \label{eqn:iterated}
\frac{E \left[ e^{\int_0^t r(s) ds} \right]}{e^{\int_0^t E[r(s)] ds} } = \frac{E_{\rho} \left[ \overbrace{E_{W}\left[ e^{\int_0^t r(s) ds} \right]}^{inner} \right] }{  e^{E_{\rho} \int_0^t E_W[r(s)] ds} }
\end{equation}
which will simplify in some cases, as we shall see, to give simple additive corrections for early stage growth arising from variation.

\subsection{Mean (inner) early stage growth}
\label{sec:inner}

\begin{figure}
    \centering
    \includegraphics[scale=0.5]{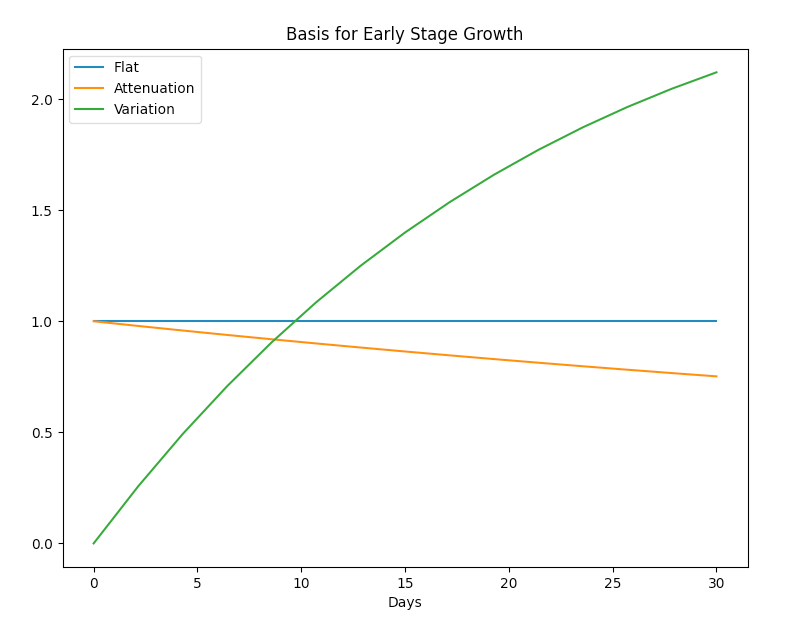}
    \caption{Basis for early stage growth shape using Equation \ref{eqn:basis}. The difference between initial infection rate and long term infection 
    rate is multiplied by the orange function $\tau/t$ which can be interpreted as including an attenuation due to repeat contacts. Variation drives a yield component shown in green, which has been scaled down by a factor of $200$ relative to the term $\frac{\tau^2}{\kappa t}$ in Equation \ref{eqn:basis}. The green line would be multiplied by $\nu_{\sigma}^2$ where  $\nu_{\sigma}  =  \frac{ \sigma }{\sqrt{2\kappa} r_{\infty}}$ is the ratio of the
    long run standard deviation of instantaneous growth to the long run mean of the same.} 
    \label{fig:basis}
\end{figure}

Unlike the deterministic mixture model, the mean growth is not obvious even for fixed $\omega$. The mean across different paths generated by $W_t$ requires an expectation of the exponential of a stochastic integral. Fixing $\omega$ for now, expected growth in infections is given by
\begin{equation}
    \label{eqn:gt}
    g_t := \frac{1}{t}\log E_{W}\left[ e^{\int_0^t r(s) ds} \right] = \frac{\tau(t) r_0(\omega) + B(t;\omega)}{t}
\end{equation}
where $\tau(t)$ and $B(t)$ are functions not depending on the initial rate $r_0(\omega)$ and we have written the expectation with a $W$ subscript to emphasize that this is with respect to the law of the stochastic process (not the probability space $\Omega$ that might do its own mixing independently). Here $r_0(\omega)$ appears only as a linear term. The function $\tau$ depends only on $\kappa_{\omega}$
$$
     \tau(t) := \frac{1-e^{-\kappa t}}{\kappa}
$$
and is approximately equal to $t$ for small times. The function $B(t;\kappa,\sigma)$ is given by 
\begin{equation}
\label{eqn:ratio}
   \frac{B(t;\omega)}{t} := \left( r_{\infty}(\omega)+\frac{\sigma_{\omega}^2}{2\kappa_{\omega}^2}
                \right) \frac{t-\tau}{t} - \frac{\sigma_{\omega}^2}{4\kappa_{\omega}^2} \frac{\tau^2}{t} 
\end{equation}
and revealed in the next section to be a thinly veiled bond price formula. The initial growth rate is $r_0(\omega)$ as we expect and the eventual growth rate is
$$
      g_{\infty}(\omega) = r_{\infty}(\omega)+\frac{\sigma_{\omega}^2}{2\kappa_{\omega}^2}
$$
although this holds only so long as $s(t)\approx 1$. We have some flexibility to match an infection rate decaying towards a lower level, but actually the $\sigma^2$ terms allow us to fit slightly more complicated term structures, as is revealed by rearranging the growth equation.
\begin{eqnarray}
\label{eqn:growth}
    g_t(\omega) & = & \overbrace{r_{\infty}(\omega) +  \frac{\sigma^2}{2\kappa^2}}^{g_{\infty}} +  \frac{\tau}{t} 
    \left\{ (r_0(\omega)-r_{\infty}(\omega)) - \frac{\sigma^2}{2\kappa^2} \left(1+\frac{\tau}{2} \right) \right\}
\end{eqnarray}
Here we might view the functions $1$, $\tau/t$ and $(1+\tau/2)(\tau/t)$ as a basis for shapes taken by growth as a function of time. Or, separating the roles of the parameters, we might introduce ratios 
\begin{eqnarray}
   \nu_0 & = & \frac{r_0-r_{\infty}}{r_{\infty}} \\
   \nu_{\sigma} & = & \frac{ \sigma }{\sqrt{2\kappa} r_{\infty}}
\end{eqnarray}
so that $\nu_{\delta}$ represents the ratio of the mean infection rate attenuation to eventual infection rate and $\nu_{\sigma}$ is the ratio of the ergodic standard deviation of the infection rate process to the long term mean. Then growth is written 
\begin{eqnarray}
\label{eqn:basis}
    \frac{g_t}{r_{\infty}} & = & 1  + \nu_{\delta} \frac{\tau}{t}  - \nu_{\sigma}^2 \frac{\tau^2}{\kappa t} 
\end{eqnarray}
We plot this basis in Figure \ref{fig:basis}, which might be viewed as a non-orthogonal basis. One might write
$$
   \frac{g_t}{r_{\infty}} =  \left[  1, \nu_0, \nu_{\sigma}^2 \right]  \left[
   \begin{array}{c}
        1  \\
        \tau/t \\
        -\frac{\tau^2}{2 \kappa t}  
    \end{array}
    \right]
$$
to emphasize a natural choice of parameters $r_{\infty}$, $\nu_0$ and $\nu_{\sigma}^2$, yield basis and resultant ratio to long term growth. 

Two of these basis functions can be accommodated in the deterministic setting also, but we can revert to that by setting $\sigma \equiv 0$ rather than treating it separately. 

\subsection{Relationship to term structure modeling}
\label{sec:bond}

Our terminology is suggestive and as noted the growth formula in Equation \ref{eqn:growth} appears elsewhere in another guise. It is the price of a bond paying no coupons in a Vasicek short rate model \cite{Vasicek1977AnStructure}. This correspondence obviates a full derivation, though that is a straightforward if lengthy exercise.\footnote{The Vasicek bond price formula is a convexity calculation for a normal random variable. Most of the terms arise from the formula for the variance of the Ornstein-Uhlenbeck process, one half of which contributes to the difference between $E[exp(\int_0^t r_s ds)]$ and $exp(E[\int_0^t r_s ds])$. } Skipping past that thanks to Vasicek's formula, we need only observe that the price of a zero coupon bond with maturity $t$, mean reversion level $x_{\infty}$, initial short rate $x_0$ and $\kappa$ and $\sigma$ as before is
$$
    B(x_0) = exp\left( E\left[-\int_0^t x_s ds\right]\right) = e^{-x_0 \tau +\Lambda(t)}
$$
where
$$
  \Lambda(t;x_0) = \left( x_{\infty} - \frac{\sigma^2}{2\kappa^2} \right)(\tau(t)-t) - \frac{\sigma^2 \tau(t)^2}{4 \kappa}
$$
and $\tau$ as above. If we let $x=-r$ with $x_0 \rightarrow -r_0$ and $x_{\infty}\rightarrow -r_{\infty}$ we have
\begin{eqnarray*}
    dr_t = -dx_t  & = & -\kappa\left( -r_{\infty} - (-r) \right) - \sigma dW_t \\
      & = & \kappa\left( r_{\infty} - r \right) + \sigma d(-W_t)
\end{eqnarray*}
and since $W_t$ and $-W_t$ have the same law, $r_t$ follows a ``Vasicek process''. Thus by using negative interest rates in this manner we have
\begin{eqnarray*}
    \overbrace{ \frac{1}{t} \log E\left[\int_0^t r_s ds\right]}^{mean\ early\ stage\ growth} & = &
    \frac{1}{t} \log \overbrace{E\left[-\int_0^t x_s ds\right]}^{Vasicek\ bond\ price} \\
    & = & \frac{1}{t} \left( -(-r_0)\tau + \Lambda\left(t;x_{\infty}=-r_{\infty}, x_0=-r_0 \right) \right) \\
    & = & \frac{1}{t} \left( r_0 \tau + 
                               \left(-r_{\infty}-\frac{\sigma^2}{2\kappa^2}\right)(\tau-t) - \frac{\sigma^2\tau^2}{4\kappa} 
                    \right) \\
    & = & \frac{1}{t}
                    \left\{ r_0 \tau + 
                               \underbrace{\left(r_{\infty}+ \frac{\sigma^2}{2\kappa^2}\right)(t-\tau) - \frac{\sigma^2\tau^2}{4\kappa}}_{B(t)} 
                    \right\}
\end{eqnarray*}
and we have justified Equation \ref{eqn:gt}.\footnote{The exponentially exploding bond price may seen counterintuitive. However a bond becomes extremely expensive when it is an attempt to shield wealth from the confiscatory effect of negative instantaneous interest rates.}

\subsection{Inner convexity}

Again let us fix $\omega \in \Omega$. From the ``inner'' mean growth rate defined in Equation \ref{eqn:gt} and given in Equation \ref{eqn:growth}, we derive the convexity adjustment defined in Equation \ref{eqn:convexity} for mean early stage growth relative to the deterministic model simply by setting $\sigma(\omega)\equiv0$. It is apparent from Equation \ref{eqn:ratio} that 
\begin{eqnarray}
     G(t) & = & \frac{1}{t} \log \left( \frac{ E_W\left[ \exp\left(\int_0^t r_s ds \right) | r(0)=r_0 \right]  }{
       \exp\left(\int_0^t E[r_s]  ds \right) } \right) \nonumber \\
       & = & B(t;\sigma)-B(t;\sigma=0) \\
       & = &  \frac{\sigma_{\omega}^2}{2\kappa_{\omega}^2}
                 \frac{t-\tau}{t} - \frac{\sigma_{\omega}^2}{4\kappa_{\omega}^2} \frac{\tau^2}{t} \nonumber \\
       & = & \frac{\sigma_{\infty}^2}{\kappa_{\omega}} \left\{ \frac{t-\tau}{t} - \frac{\tau^2}{t}        \right\} \rightarrow \frac{\sigma_{\infty}^2}{\kappa_{\omega}} 
\end{eqnarray}
where we recall that 
$$
     \sigma_{\infty}^2 = \frac{\sigma^2}{2\kappa}(\omega)
$$
is the long term ergodic variance of the infection rate $\beta$. This is the multiplicative adjustment that needs to be made to account for variation induced by different paths taken by infection in different places, as compared with simply assuming infection rate is the same everywhere.

\subsection{Overall convexity}

Referring to Equation \ref{eqn:iterated} we next consider the overall convexity taking into account variation due to $\omega \in \Omega$, and show that this is easily computed for some choices. In other words we allow $\eta$ to determining the law of $\beta_0=\bb\eta$ in analogy with the deterministic SIR case considered in Section \ref{sec:definetti}. Using an expression for the mean exponentialed integral analogous to Equation \ref{eqn:growth}, we have

\begin{eqnarray}
\label{eqn:outer}
    G(t) & = & \frac{1}{t} \log \frac{
E_\rho\left[ \exp\left( 
r_{\infty}(\omega) + 
\frac{\sigma^2}{2\kappa^2} +  
\tau  \left\{ (r_0(\omega)-r_{\infty}(\omega)) - \frac{\sigma^2}{2\kappa^2} \left(1+\frac{\tau}{2} \right) \right\}
\right) \right]}{
E_\rho\left[ \exp\left( 
r_{\infty}(\omega) +  
\tau   (r_0(\omega)-r_{\infty}(\omega))  
\right) \right]
} 
\end{eqnarray}
This will simplify in some cases we now consider.

\subsection{Independent variation}
Suppose now that under the measure $\rho$ the variables $r_{\infty}$, $\sigma$, $\kappa$ and $r_0$ vary independently. The numerator in Equation \ref{eqn:outer} then factors into product terms, many of which cancel immediately with terms on the denominator. 
\begin{eqnarray}
\label{eqn:outerindependent}
    G(t) & = & \frac{1}{t} \log \frac{
    E_\rho\left[ \exp\left( 
r_{\infty}(\omega)t +  
\tau   (r_0(\omega)-r_{\infty}(\omega))  
\right) \right]
E_\rho\left[ \exp\left( t
\frac{\sigma^2}{2\kappa^2} +  
\tau  \left\{ - \frac{\sigma^2}{2\kappa^2} \left(1+\frac{\tau}{2} \right) \right\}
\right) \right]}{
E_\rho\left[ \exp\left( 
r_{\infty}(\omega)t +  
\tau   (r_0(\omega)-r_{\infty}(\omega))  
\right) \right]
} \nonumber \\
& = & 
\frac{1}{t} \log 
E_\rho\left[ \exp\left( 
t\frac{\sigma^2}{2\kappa^2} +  
\tau  \left\{ - \frac{\sigma^2}{2\kappa^2} \left(1+\frac{\tau}{2} \right) \right\}
\right) \right] \nonumber \\
& = &
\frac{1}{t} \log 
E_\rho\left[ \exp\left( 
t \frac{\sigma^2}{2\kappa^2} \right) \right]  +
\frac{1}{t} \log E_\rho\left[ \exp\left( 
-\tau  \frac{\sigma^2}{2\kappa^2} \left(1+\frac{\tau}{2} \right) 
\right) \right]
\end{eqnarray}
which we recognize as a sum of convexity adjustments that might be independently computed. For example the first term is a convexity adjustment pertaining to the ergodic mean variance. If we write 
$$
  \sigma^2_{\infty}(\omega) = \sigma^2_{\infty}\eta(\omega)
$$
for some convenient choice of variation function $\eta$ then the approach of Section \ref{sec:convexity} applies (assuming independence allows this to be recycled in the second term also). 

\subsection{Additive convexity adjustments for growth}

To focus on this first term in Equation \ref{eqn:outerindependent} momentarily, which is the only one to survive several mean generation times:
$$
\frac{1}{t}  \log E_\rho\left[ \exp\left( t
\frac{\sigma^2}{2\kappa^2} \right) \right] = 
\frac{1}{t}  \log \exp\left( t E\left[ 
\frac{\sigma^2}{2\kappa^2} \right] G_{\Omega} \right) =  
G_{\Omega} + G_W 
$$
where $G_{\Omega}$ is a convexity adjustment arising from variation of $\sigma^2_{\infty}$ over $\Omega$. On the other hand $G_W$ might also be called a convexity adjustment although it arises from the {\em existence} of $\sigma^2_{\infty}$ and is proportional to the mean value of the ergodic variance. 
$$
   G_W = E_{\rho} \left[ \frac{\sigma^2}{2\kappa^2} \right]
$$
In this way a simple meta-population model with two types of variation is easy to relate to its deterministic counterpart at least as far as early stage growth is concerned. This property arises from choice of an infection rate falling into the affine class, as we remarked earlier, which is evidently a tractable choice. In contrast, if early stage growth is not linear in the initial value of $\beta_0$ then we have more work to do. Likewise if a complicated dependency is assumed between the variables $r_{\infty}$, $\sigma$, $\kappa$ and $r_0$ as functions of $\omega$ then no simple calculation such as this may apply. This is not to suggest that there are not other ways to make progress simplifying Equation \ref{eqn:growth} for other choices.   

\subsection{Peak infection}

\label{sec:meanvar}
\label{sec:h}

The instantaneous infection rate at the onset of herd immunity will determine whether balance occurs between recovery and new infections (as compared with the mean growth up until that point in time).\footnote{There is a tendency to conflate these two quantities, and the corresponding $R$ values, in a constant parameter compartmental model.} The mean and variance of $\beta(t)$ are 
\begin{eqnarray}
     E[ \beta(t) ] & = &  \beta_{\infty} + e^{-\kappa t} \left(\beta_0 - \beta_{\infty} \right) \\
     var(\beta(t)) & = &  \frac{\sigma^2}{2\kappa}\left(1-e^{-2\kappa t} \right)
\end{eqnarray}
so with
$$
      \sigma_{\infty} = \frac{\sigma}{\sqrt{2\kappa}}
$$
as shorthand then the ergodic distribution is $\beta(t \rightarrow \infty) \sim N( \beta_{\infty}, \sigma^2_{\infty} )$. As noted in Equation \ref{eqn:peakinfection} an approximate convexity adjustment for peak infection might break down into two components: one arising from deviation of the harmonic mean from the arithmetic mean and the other arising from a superposition effect.

Some approximations for peak infection might be couched in terms of an auxiliary function 
\begin{eqnarray}
\label{eqn:h}
    h(a,b,c; \mu,\sigma) := \frac{1}{E\left[\frac{1}{a+b \max(X,c)} | X \sim N(\mu,\sigma^2) \right]}
\end{eqnarray}
which is a generalization of the harmonic mean. To see this, note that the susceptible population can be approximated as 
$$
    s(\omega,t) := E_W\left[ \frac{\gamma}{max(\beta(t),\gamma)} \right] = E_W\left[ \frac{1}{max(\beta(t)/\gamma,1)} \right] = h(1,0,1; \mu,\sigma)
$$
where
$$\mu=\frac{ \beta_{\infty}+e^{-\kappa t}(\beta_0-\beta_{\infty}) 
                }{\gamma}
    $$
    and 
    $$\sigma=\sigma_{\infty}(1-e^{-2\kappa t})/\gamma$$ are the mean and variance of the infection rate. 
This approximation assumes that when $\beta(t)<\gamma$ the epidemic never really gets going and the susceptible population remains close to unity. 

\subsection{Harmonic mean of gaussian}
\label{sec:harmonictaylor}
Using the Taylor expansion
$$
  \frac{1}{1+x} \approx 1 - x + x^2 - x^3 + x^4 + O(x^5)
$$
valid for $x \in (0,1)$ let $\{\epsilon_i\}_{i=1}^n$ be independent standard normal. In the limit $n \rightarrow \infty$ we have  
\begin{eqnarray*}
   h(0,1,\infty;\mu,\sigma) & = & \frac{1}{E\left[\frac{1}{X} | X \sim N(\mu,\sigma^2) \right]} 
    =  \frac{1}{E\left[\frac{1}{\mu+\sigma Z} | Z \sim N(0,1) \right]} \\
      & \approx & \frac{n}{\sum_{i=1}^n \frac{1}{\mu+\sigma \epsilon_i}} \\
      & \approx & \frac{n}{ \frac{1}{\mu} \sum_{i=1}^n \left( 
            1 - \frac{\sigma}{\mu}\epsilon_i +  \frac{\sigma^2}{\mu^2}\epsilon_i^2
            -   \frac{\sigma^3}{\mu^3} \epsilon_i^3 + \dots
      \right) } \\
      & = & \frac{\mu}{
            1- \frac{\sigma}{\mu} \frac{1}{n}\sum_{i=1}^n \epsilon_i + 
            \frac{1}{n}\sum_{i=1}^n  \frac{\sigma^2}{\mu^2}\epsilon_i^2
            - \frac{1}{n}\sum_{i=1}^n    \frac{\sigma^3}{\mu^3} \epsilon_i^3 
      } \\
     & \approx & 
     \mu \left( 
            1 + \frac{\sigma}{\mu} \frac{1}{n}\sum_{i=1}^n \epsilon_i -
            \frac{1}{n}\sum_{i=1}^n  \frac{\sigma^2}{\mu^2}\epsilon_i^2
            + \frac{1}{n}\sum_{i=1}^n    \frac{\sigma^3}{\mu^3} \epsilon_i^3 
       + \dots \right) \\
      & \rightarrow & 
         \mu\left( 1 -  \frac{\sigma^2}{\mu^2} - 3  \frac{\sigma^4}{\mu^4} - \dots \right)
\end{eqnarray*}
because the odd powers have zero mean and the even powers are the well known moments of the normal distribution. The simplest convexity adjustment for peak infection using only the first term is 
\begin{eqnarray}
\label{eqn:vasicekj}
    J & = &   \left( 1 -  \frac{\sigma^2}{\mu^2}\right)^{-1} \nonumber \\
      & \approx &   1 +  \frac{\sigma_{\infty}(1-e^{-2\kappa t})/\gamma}{\left( \frac{ \beta_{\infty}+e^{-\kappa t}(\beta_0-\beta_{\infty}) 
                }{\gamma}\right)^2} \nonumber \\
     & = & 1 +   \frac{\gamma \sigma_{\infty}(1-e^{-2\kappa t})}{\left( \beta_{\infty}+e^{-\kappa t}(\beta_0-\beta_{\infty}) 
                \right)^2} \nonumber \\
    & \rightarrow & 1 +   \frac{\gamma}{\beta_{\infty}} 
                                    \frac{\sigma_{\infty}}{\beta_{\infty}}
\end{eqnarray}
where we recognize the two ratios. 

\subsection{A simple adjustment to peak infection susceptibles}

The adjustment in Equation \ref{eqn:vasicekj} is an easy mental calculation as $R_{\infty}:=\beta_{\infty}/\gamma$ is a kind of limiting $R_0$ value (to mix conventions somewhat) whereas $\sigma_{\infty}/\beta_{\infty}$ is the limiting {\em relative} variation of infection. Thus to translate relative variation into a herd immunity adjustment we can use the very coarse but easy to remember formula:
$$
       susceptible\ multiplier = 1 + \frac{relative\ variation}{R}
$$
This, we emphasize, is the number we should multiply the susceptible population by to correct an estimate that is derived by assuming infection follows a deterministic trajectory. Note that even for $R$ values such as $R=2$ or $R=4$ this is an appreciable correction. We can also use more accurate approximations using the derivation of Equation \ref{eqn:vasicekj}, of course.

%% file: tractable.tex
\section{Appendix A: Additional convexity adjustments}

We provide further examples of convexity adjustments for peak infection, early stage growth and population profile. 

\subsection{Peak infection}

Section \ref{sec:shape} considers examples where $\eta$ depends on population density $\rho$. Here are some example where it does not. 

\subsubsection{Inverse Gamma}

We suppose the law of $\eta$ is an Inverse Gamma distribution with parameters $\alpha>1$ and $\beta$ chosen such that the mean $\frac{\beta}{\alpha-1}=1$. That is to say $\beta=\alpha-1$. The mean of the inverse is $\alpha/\beta$ and thus
$$
     J(\alpha) = E\left[\frac{1}{\eta} \right] = \frac{\alpha}{\alpha-1}
$$
In passing, we note that this example might be formalized by letting the probability space be $\Omega=(0,1)$ with uniform $\rho$ equivalent to Lebesgue measure and $\eta$ the inverse cumulative distribution function for the Inverse Gamma function. In similar fashion other examples can be made to satisfy the definition we provided for the variation function $\eta$, although when $\eta$ does not depend on $\rho$ this may seem a tad ceremonial.  

\subsubsection{Lognormal}

Suppose instead the law of $\eta$ is log-normally distributed, which is to say that with mild abuse of notation
$$
    \eta \sim  e^{\mu+\sigma Z}
$$
for standard normal $Z$. The mean is
$$ 
    E[\eta] = e^{\mu+\frac{1}{2}\sigma^2}
$$
whereas the harmonic mean is 
$$
   H(\mu) = e^{\mu-\frac{1}{2}\sigma^2} 
$$
The ratio $J$ is the ratio of the mean to the harmonic mean and thus 
$$
     J(\sigma) = e^{\sigma^2}
$$
We set $\sigma^2=-2\mu$  so that the mean is $1$. So alternatively we can write
$$
    J(\mu) = e^{-2\mu}
$$
bearing in mind $\mu<0$.

\subsubsection{Pareto}

Fix $\alpha>1$ and let $\sigma=1-\frac{1}{\alpha}$. Assume the law of $\eta$ follows a Pareto distribution, which is to say that the probability it exceeds $x$ is $\left( \frac{x}{\sigma} \right)^{\alpha}$ for $x>\sigma$. We have
$$
    J =  \overbrace{ \frac{\alpha \sigma}{\alpha-1} }^{mean (=1)}      / \overbrace{ \sigma \left(1+1/\alpha \right) }^{harmonic\ mean} = \frac{1}{1-\frac{1}{\sigma^2}}
$$

\subsection{Alternative harmonic mean of gaussian}
\label{sec:harmonicga}

The following approximation of the harmonic mean of a gaussian not centered near zero is provided by Dimitri Offengenden. One starts with the following approximation
$$
  \frac{1}{1+x} \approx 1 - x + \frac{3}{4}x^2 - x^3 
$$
which Offengenden finds is approximately valid for $x \in (0,1)$ with error no more than one to two percent, and at the same time quite accurate for larger values of $x$ (fitting considerably better than the Taylor expansion, for example). Let $\{\epsilon_i\}_{i=1}^n$ be independent standard normal. In the limit $n \rightarrow \infty$ we have  
\begin{eqnarray*}
   h(0,1,\infty;\mu,\sigma) & = & \frac{1}{E\left[\frac{1}{X} | X \sim N(\mu,\sigma^2) \right]} 
     =  \frac{1}{E\left[\frac{1}{\mu+\sigma Z} | Z \sim N(0,1) \right]} \\
     & \approx & 
     \mu \left( 
            1 + \frac{\sigma}{\mu} \frac{1}{n}\sum_{i=1}^n \epsilon_i -
            \frac{1}{n}\sum_{i=1}^n \frac{3}{4}\frac{\sigma^2}{\mu^2}\epsilon_i^2
            + \frac{1}{n}\sum_{i=1}^n \frac{1}{4} \frac{\sigma^3}{\mu^3} \epsilon_i^3 
       \right) \\
      & \rightarrow & 
         \mu\left( 1 - \frac{3}{4}\frac{\sigma^2}{\mu^2} \right)
\end{eqnarray*}
where $\epsilon_i$ are independent standard normal.

\subsection{Peak infection with modified harmonic mean}

In Section \ref{sec:h} we considered an approximation using a modified harmonic mean $h$. There are 
no doubt many ways to approximate this. Here we make two observations only
\begin{enumerate}
    \item It suffices to consider the case of standard normal
    \item Option pricing technology can be used
\end{enumerate}
Recall the auxiliary function that generalizes the harmonic mean:
\begin{eqnarray*}
    h(a,b,c; \mu,\sigma) := \frac{1}{E\left[\frac{1}{a+b \max(X,c)} | X \sim N(\mu,\sigma^2) \right]}
\end{eqnarray*}
Let $Y:=\frac{X-\mu_X}{\sigma_X}$ be an affine transformed version of $X$. As $Y$ is manifestly a standard normal random variable, and since
\begin{eqnarray*}
     \max(X,1) & = & \max\left( Y\sigma_X+\mu_X, 1 \right) \\
            & = & \mu_X + \max\left( Y\sigma_X, 1 -\mu_X \right)  \\
            & = & \mu_X + \sigma_X \max \left( Y, \frac{1-\mu_X}{\sigma_X} \right)
\end{eqnarray*}
it is clear that 
$$
      h(1,0,1,\mu,\sigma) = E\left[ \frac{1}{\mu_X + \sigma_X \max \left( Y, \frac{1-\mu_X}{\sigma_X} \right)} \right] = h\left( \mu_x, \sigma_x, \frac{1-\mu_x}{\sigma_x};0,1 \right)
$$
allowing one to specialize to the case of standard normal random variables when approximating $h$. Furthermore
\begin{eqnarray}
      h(a,b,c;0,1) & = & \frac{1}{2\pi} \int_{-\infty}^{\infty} \frac{e^{-\frac{1}{2}x^2}} {a+b \max(x,c)} dx \\
      & = & \frac{1}{2\pi} \frac{1}{a+bc} \int_{-\infty}^c e^{-\frac{1}{2}x^2} dx
      + \frac{1}{2\pi} \int_{c}^{\infty} \frac{e^{-\frac{1}{2}x^2}} {a+bx} dx
\end{eqnarray}
where the first term is merely a multiple of the cumulative normal distribution function. In Section \ref{sec:harmonicg} we considered the case $c=-\infty$ where only the second term survives. The technique used there to approximate $\frac{1}{a+bx}=\frac{1}{a}\frac{1}{1+\frac{b}{a}x}$ could be applied to convert the second term into a sum of normal options.\footnote{We again thank Dimitri Offengenden for the suggestion.} This might be overkill, admittedly, and the reader likely won't need reminding that this class of approximations assumes independent equilibria are reached between populations - likely an assumption leading to greater error.\footnote{In future work we hope to use these convexity adjustments as features for predicting actual peak infection levels in agent simulations, thereby providing some calibration useful for actual epidemics.} 

\subsection{Early stage growth}
\label{sec:tractablegrowth}

In order to provide more tractable examples of early stage growth adjustments due to variation, it obviously behoves us to consider variation functions with known moment generating functions. These examples are standard. We merely enforce normalization, which sometimes reveal bounds on the growth convexity adjustments. Recall that we use
$$
      t' = \bb t
$$
to measure time in units that are convenient for this purpose.

\subsubsection{Gamma}

Let $\Omega = (0,1)$ with uniform measure $\rho$ and define $\eta$ by the inverse cumulative distribution function of the Gamma distribution. Then the law of $\eta$ is Gamma distributed and 
\begin{equation}
\label{eqn:gamma}
       E[e^{t' \eta}] = (1- t \theta)^{-k}
\end{equation}
where $k$ and $\theta$ are the Gamma distribution parameters. We must set $k=1/\theta$ to achieve a mean $E[\eta]=1$. The probability density function is 
$$
   \frac{1}{\Gamma(k)\theta^k} x^{k-1} e^{-x/\theta}
$$
but will only serve us if $\theta$ is small enough that $t'<1/\theta$, for otherwise Equation \ref{eqn:gamma} is invalid. If it is valid then 
$$
       G(t) = -k \frac{log(1-\bb t/k)}{\bb t} \approx 1 +\frac{1}{2}\theta \bb t < \frac{3}{2}
$$
where the inequality follows from the limitation $t'<1/\theta$ corresponding to $t<\frac{1}{\bb \theta}$. 

\subsubsection{Poisson}

Similarly we can allow $\eta$ to be Poisson distributed with parameter $\lambda=1$. Then 
\begin{equation}
\label{eqn:poisson}
       E[e^{t' \eta}] = e^{e^{t'}-1}
\end{equation}
and thus 
$$
      G(t) = \frac{e^{\bb t}-1}{\bb t} \approx 1 + \frac{1}{2} \bb t + \frac{1}{6}\bb^2 t^2 + O(t^3) 
$$
Excess growth is unbounded. However a more careful analysis would curtail growth for some values of $\eta$ where the approximation $s(t)\approx 1$ is no longer valid. 

\subsubsection{Negative binomial}

Let $\eta$ be distributed as the the negative binomial distribution with parameters $r$ and $p$, and with $p=\frac{1}{1+r}$ so that $E[\eta]=1$. Then 
\begin{equation}
\label{eqn:negativebinomial}
       E[e^{t' \eta}] = \frac{(1-p)^r}{(1-pe^{t'})^r}
\end{equation}
and so 
$$
      G(t) = r \frac{ \log \frac{1-p}{1-pe^{\bb t}}}{\bb t} 
$$
is the excess growth. 

\subsubsection{Binomial}

With $\eta$ Binomial(n,p) and $p=1/n$ a similar calculation yields
$$
     G(t) = \frac{ \log( 1-p+pe^{\bb t})}{\bb t} \approx 1 + \frac{1}{2} t \bb (1-p) + \frac{1}{6} \bb^2 (2 p^2 - 3p + 1) t^2 + O(t^3)
$$

\subsubsection{Linear}

Let $\Omega=(0,2)$ with $\eta(\omega)=\omega$ and $\rho(\omega)=\frac{1}{2}$ so that $E[\eta]=1$ as required. Since
$$
     E[ e^{t \bar{\beta}\eta} ]  = \frac{e^{2 \bb t}-1}{2 \bb t} 
$$
we have 
$$
       G(t) =  \frac{1}{\bb t} \log\left(  \frac{e^{2 \bar{\beta}t}-1}{2 \bar{\beta} t}   \right) \approx 1 + \frac{1}{6} \bar{\beta} t 
$$
in agreement with the previous gaussian approximation for small $t$ after we note that the variance of $\eta$ is $\sigma^2=1/3$. But when $t$ is several multiples of of the timescale $1/\beta$ the limit $t \gg 1/\bar{\beta}$ becomes more relevant. We have
$$
      G(t) \rightarrow 2 
$$

\subsubsection{Quadratic}

In similar fashion we could quadratic $\eta$. We set $\Omega=(0,\sqrt{3})$, $\rho = \frac{1}{\sqrt{3}}$ and $\eta(\omega)=\omega^2$. Then $E[\eta]=1$ and $E[(\eta-1)^2]=\frac{4}{5}$. We find
\begin{eqnarray*}
    E[e^{t \bar{\beta} \eta}] & = & \frac{1}{\sqrt{ 3 }} \int_0^{\sqrt{3} } e^{\bar{\beta} t \omega^2} d\omega 
\end{eqnarray*}
Using a series expansion at $\bar{\beta}t \rightarrow \infty$ we approximately have
$$
   \frac{ \log  E[e^{t \bar{\beta} \eta}] }{t} \rightarrow \frac{1}{t} \log \left( e^{3\bar{\beta}t} \left\{ \frac{1}{\bar{\beta}t} + \frac{1}{36 \bar{\beta}^2 t^2 } \right\} \right) \rightarrow 3 \bar{\beta} 
$$
showing that if the limit is relevant, the growth inflated by recovery is three times what it would be $\beta$ were to be held constant. 
$$
    G(t) \rightarrow  3 
$$

\subsection{Additional population profile adjustments}

\label{sec:tractableshape}

Following on from Section \ref{sec:shape} we provide more examples of convexity adjustments when contagion varies as a power law of population.

\subsection{Linear drop in population density}

\begin{figure}
    \centering
    \includegraphics[scale=0.45]{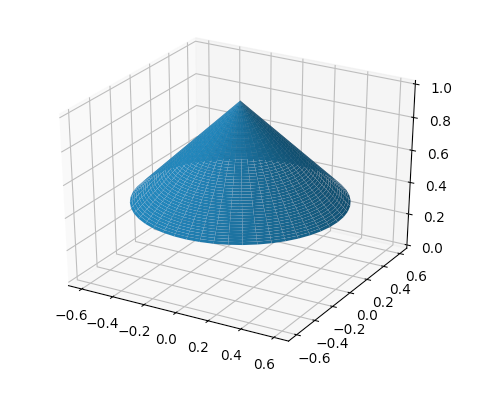}
    \caption{An example of a stylized population density profile resembling a circus tent.}
    \label{fig:tentpop}
\end{figure}

The city's population takes the form of a circus tent shown in Figure \ref{fig:tentpop}. 

\begin{figure}
    \centering
    \includegraphics[scale=0.75]{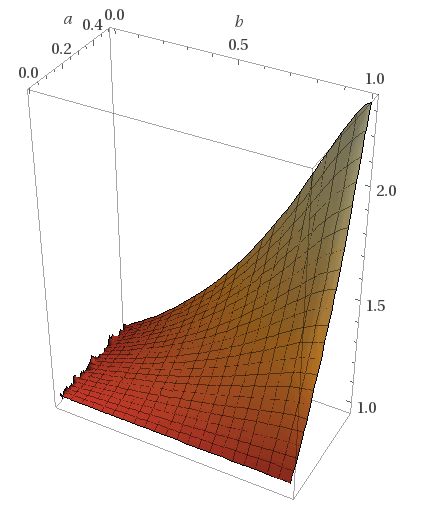}
    \caption{Convexity adjustment to the susceptible population as a function of parameters $a$ and $b$ determining contagiousness in a tent-shaped population profile. Population density falls off linearly as we move away from the origin until we reach a radius $b<1$ as shown in Figure \ref{fig:tentpop}. Contagiousness is related to density by a power law with coefficient $a \ll 1$. Even a weak relationship between population density and contagiousness can imply a significant ratio between the number of susceptible people at peak infection and the number of susceptible people at peak infection when we assume homogeneous density.}
    \label{fig:tent}
\end{figure}

Here we assume the city has a radius of $b$ and the tent's roof, pitched at a height $1$, falls off at a 45 degree angle. Quite unlike a circus tent it is assumed that the height of this roof represents population density, which drops abruptly to zero as we reach a distance $b$ from the center of town.\footnote{That is to say that all citizens reside in a conical version of the Luxor Hotel.} A calculation analogous to those above yields
$$
J(a,b) = \frac{ \frac{1}{(a^2-5a+6)} - \frac{ (1-b)^{2-a}( 1+b(2-a) ) }{  (2-a)(3-a) } }
         { \frac{1}{a^2+5a+6} -  \frac{ (1-b)^(a+2)( (a+2)*b + 1 ) }{  (2+a)(3+a)} }
$$
The multiplicative correction takes values on the order of $1.5$ for $a \approx 1/4$ and $b \approx 0.9$, for example. This function is plotted in Figure \ref{fig:tent} 

\subsubsection{Power law population distribution}

An example when $J$ is unbounded is provided by a ``skirt''. Suppose $\rho(r) \propto 1/r^3$ and the population lives outside the unit circle. Then
$$
   J(\alpha)  =   \frac{  \int_1^{\infty} \frac{1}{r^3(1-\alpha)}   r dr  }
                    {  \int_1^{\infty} \frac{1}{r^3(1+\alpha)}   r dr } = \frac{1+3a}{1-3a} \rightarrow \infty 
$$
It is of note that the correction is inaccurate beyond small values of $\alpha$. The size of the immune group cannot exceed unity and this is not taken into account.  

\subsection{Gamma distributed populations}
With density proportional to $re^{-r}$ and $\beta(r)$ the $\alpha$ power of the same, 
\begin{eqnarray*}
J_1(\alpha) & = & \frac{  \int_0^{\infty} (r exp(-r))^{1+\alpha} r dr  }
            {  \int_0^{\infty} (r exp(-r))^{1-\alpha} r dr  }\\
        & = & \frac{ (1 + \alpha)^{-\alpha-3} \Gamma(3 + \alpha) }
                   { (1 - \alpha)^{\alpha-3} \Gamma(3 - \alpha)} \\
       & \approx &  \left( 3/2+\gamma \alpha \right)^2 + O(\alpha^3) \\
       & \approx & 1 + 4.15 \alpha
\end{eqnarray*}
where $\gamma$ is the Euler-Mascheroni constant, $\gamma \approx 0.57721$. If instead density is proportional to $r^2 e^{-r}$ then
\begin{eqnarray*}
J_2(\alpha) & = & \frac{  \int_0^{\infty} (r^2 exp(-r))^{1+\alpha} r dr  }
            {  \int_0^{\infty} (r^2 exp(-r))^{1-\alpha} r dr  }\\
        & = & \frac{ (1 + \alpha)^{-4-2a} \Gamma(4 + 2 \alpha) }
                   { (1 - \alpha)^{2a-4}  \Gamma(4 - 2 \alpha) } \\
       & \approx &  \left( 1 - (1/3+2\gamma) \alpha \right)^2 + O(\alpha^3) \\
       & \approx & 1 + 2.97 \alpha
\end{eqnarray*}
Notice that these adjustment are material even if the power law $\alpha$ relating population density to contagiousness is relatively small, say $\alpha \approx 0.1$.

%% file: physical.tex
\section{Appendix B: Justifying decaying mean-reverting infection rate}
\label{sec:justifyvasicek}

We provide further motivation for the choice of Ornstein-Uhlenbeck infection rates

\subsection{Form suggested on phenomenological basis}

Chowell et al \cite{Chowell2016MathematicalReview} propose modeling infection rates as time varying:
\begin{equation}
\label{eqn:chowell}
    \beta(t) =  \beta_0 \left\{ (1-\phi)e^{-qt}+\phi \right\}
\end{equation}
We refer the reader to the extensive empirical survey by these authors that led them to suggest this model for early stage growth. The reader will recognize this as the mean of an Ornstein-Uhlenbeck process. 

\subsection{Repeat contact in a continuous physical model}

\begin{figure}
    \centering
    \includegraphics[scale=0.35]{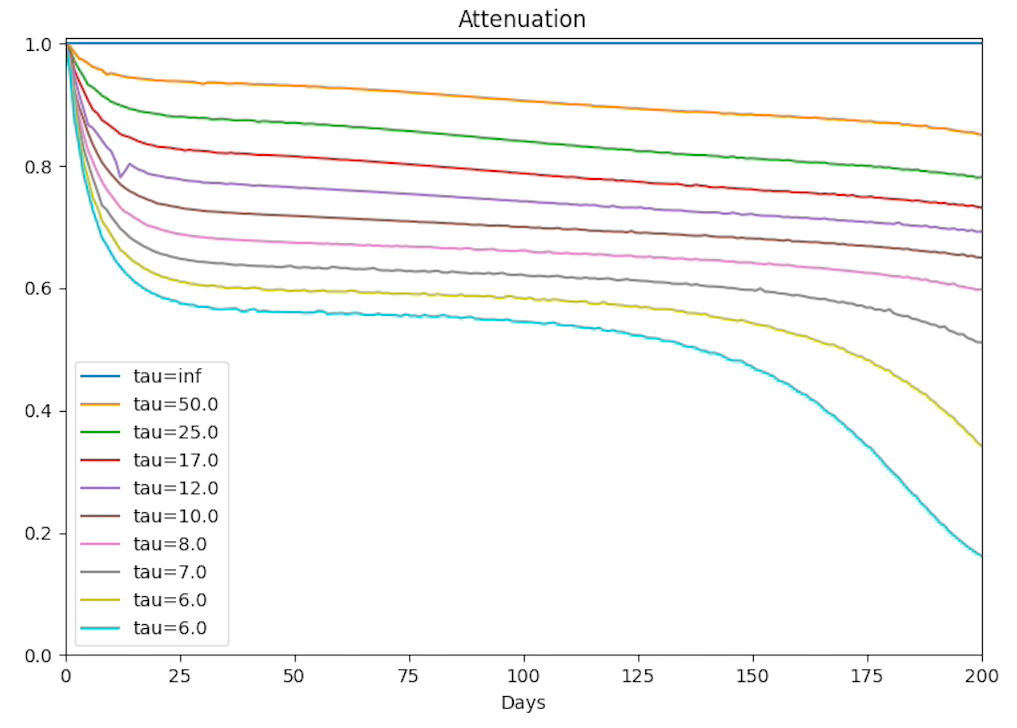}
    \caption{A physical infinite agents model provides further motivation for choice of decaying mean-revering infection rate. In \cite{Cotton2020RepeatDisease} the variation in infection rate $\beta(t)$ revealed by numerical solution of delay differential equations in a modified SIR compartmental model. The form taken accounts for repeat contacts. It can be seen that infection rate is reasonably well approximated by an Ornstein-Uhlenbeck process - at least for early stage growth and up until the time of peak infection - the period relevant to convexity adjustments presented here (in the late stage of the epidemic a shifting in the vintage of the infected cohort may lead to a second drop in infection rate).}
    \label{fig:attenuation}
\end{figure}

In \cite{Cotton2020RepeatDisease} the probability of novel contact in an infinite agents model is shown to take the form
$$
    \beta(t;\beta_0, \alpha_0) =  \beta_0 \overbrace{\left(\frac{1-e^{-\alpha_0 t}}{\alpha_0 t}\right)}^{Ein(\alpha_0 t)}
$$
This model assumes that agents populate the plane $\mathcal{R}^2$ and make visits to nearby locations with probability proportional to a bivariate gaussian distribution. When probability of repeat contact is taken into account we can approximate the system by a compartmental model with delay differential equation for infection. In this approach the infection rate is multiplied by an average over the infected cohort 
$$
      \bar{P}(t;\alpha) =  \frac{ \int_{s=0}^t Ein(\alpha(t-s)) \overbrace{\left[ \frac{\partial i(s)}{\partial s} +\gamma i(s) \right]}^{new\ infections} e^{-\gamma (t-s)} ds}{\int_{s=0}^t \left[ \frac{\partial i(s)}{\partial s} +\gamma i(s) \right] e^{-\gamma (t-s)} ds } 
$$
Because this is a vintage effect it may be seen to fall within a broad class of models considered by Kermack and McKendrick \cite{1927AEpidemics}, \cite{Kermack1991ContributionsEndemicity} \cite{Kermack1991ContributionsEndemicityb}. This particular choice leads to infection rates resembling the patterns seen in Figure \ref{fig:attenuation}, which can be seen to be reasonably well approximated by our choice. The reader is referred to the paper for further details \cite{Cotton2020RepeatDisease}.
